\documentclass[10pt,aps,pra,reprint,superscriptaddress]{revtex4-1}

\usepackage[colorlinks=true,linkcolor=blue,citecolor=blue,urlcolor=blue]{hyperref}
\usepackage{amssymb}
\usepackage{graphicx}
\usepackage[exponent-product=\cdot,per-mode=symbol,range-phrase=--,range-units=single]{siunitx}

\makeatother
\begin{document}

\title{Evaluation of ambient dose equivalent rates influenced by vertical and horizontal distribution of radioactive cesium in soil in Fukushima Prefecture}

\author{Alex Malins}\email{malins.alex@jaea.go.jp}
\affiliation{Center for Computational Science \& e-Systems, Japan Atomic Energy Agency, 178-4-4 Wakashiba, Kashiwa, Chiba, 277-0871, Japan}

\author{Hiroshi Kurikami}
\affiliation{Sector of Fukushima Research and Development, Japan Atomic Energy Agency, 4-33 Muramatsu, Tokai-mura, Naka-gun, Ibaraki, 319-1194, Japan}

\author{Shigeo Nakama}
\affiliation{Sector of Fukushima Research and Development, Japan Atomic Energy Agency, 1-29 Okitama-cho, Fukushima-shi, Fukushima, 960-8034, Japan}

\author{Tatsuo Saito}
\affiliation{Sector of Fukushima Research and Development, Japan Atomic Energy Agency, 4-33 Muramatsu, Tokai-mura, Naka-gun, Ibaraki, 319-1194, Japan}

\author{Masahiko Okumura}
\affiliation{Center for Computational Science \& e-Systems, Japan Atomic Energy Agency, 178-4-4 Wakashiba, Kashiwa, Chiba, 277-0871, Japan}

\author{Masahiko Machida}
\affiliation{Center for Computational Science \& e-Systems, Japan Atomic Energy Agency, 178-4-4 Wakashiba, Kashiwa, Chiba, 277-0871, Japan}

\author{Akihiro Kitamura}
\affiliation{Sector of Fukushima Research and Development, Japan Atomic Energy Agency, 4-33 Muramatsu, Tokai-mura, Naka-gun, Ibaraki, 319-1194, Japan}

\date{\today}

\begin{abstract}
The air dose rate in an environment contaminated with \textsuperscript{134}Cs and \textsuperscript{137}Cs depends on the amount, depth profile and horizontal distribution of these contaminants within the ground. This paper introduces and verifies a tool that models these variables and calculates ambient dose equivalent rates at \SI{1}{\metre} above the ground. Good correlation is found between predicted dose rates and dose rates measured with survey meters in Fukushima Prefecture in areas contaminated with radiocesium from the Fukushima Dai-ichi Nuclear Power Plant accident. This finding is insensitive to the choice for modelling the activity depth distribution in the ground using activity measurements of collected soil layers, or by using exponential and hyperbolic secant fits to the measurement data. Better predictions are obtained by modelling the horizontal distribution of radioactive cesium across an area if multiple soil samples are available, as opposed to assuming a spatially homogeneous contamination distribution. Reductions seen in air dose rates above flat, undisturbed fields in Fukushima Prefecture are consistent with decrement by radioactive decay and downward migration of cesium into soil. Analysis of remediation strategies for farmland soils confirmed that topsoil removal and interchanging a topsoil layer with a subsoil layer result in similar reductions in the air dose rate. These two strategies are more effective than reverse tillage to invert and mix the topsoil.
\end{abstract}

\maketitle

\section{\label{sec:Introduction}Introduction}

Dose reconstruction performed after the accident at the Fukushima Dai-ichi Nuclear Power Plant (FDNPP) showed that in the main regions affected by the accident, like the evacuated areas, external exposure to radiation from radionuclides deposited on the ground (groundshine) was the most important pathway contributing to effective doses~\citep{WHO2012,UNSCEAR2014}. Since the accident the Japanese Government has restricted the sale of contaminated foodstuffs, and the short-lived tellurium, iodine and xenon radioisotopes released (\textsuperscript{131m}Te, \textsuperscript{132}Te, \textsuperscript{131}I, \textsuperscript{132}I, \textsuperscript{133}I, \textsuperscript{133}Xe) have decayed to completion. Therefore, the main radiological hazard that persists in the environment is exposure to groundshine from radioactive cesium (\textsuperscript{134}Cs and \textsuperscript{137}Cs).

Groundshine after a nuclear accident tends to decrease due to radioactive decay of short-lived isotopes (e.g.~\textsuperscript{132}Te, \textsuperscript{131}I, \textsuperscript{132}I and \textsuperscript{134}Cs) and the penetration of fallout radionuclides into soil~\citep{ICRU53}. The Japan Atomic Energy Agency (JAEA) and partner organizations have been monitoring the environment in North-East Japan since the accident in March 2011 under contract from the Japanese Government. In particular, the consortium has been measuring radiocesium activity depth distributions within soil and monitoring air dose rates at locations of flat, undisturbed fields~\citep{Saito2015}.

Understanding the relationship between distributions of radioactive cesium within the ground and air dose rates is vital for tracking radiocesium migration, predicting future dose rates, and evaluating remediation strategies for reducing dose rates. Previous authors have published conversion factors between the concentration of radionuclides within the ground and various air dose rate quantities~\citep{Beck1968,Beck1972,Beck1980,Saito1995,Quindos2004,Saito2014,Askri2015}. These conversion factors assume spatially constant radionuclide inventories and depth distributions. To assist the recovery from the Fukushima disaster, \citet{Satoh2014} developed a calculation system to evaluate air dose rates allowing spatially varying radionuclide inventories, using a method based on summing contributions from radionuclides in different volumes of soil.

In this paper we present and verify a tool to calculate ambient dose equivalent rates to high precision for arbitrary depth profiles and horizontal distributions of \textsuperscript{134}Cs and \textsuperscript{137}Cs fallout within soil. We describe the workings of the tool and demonstrate the validity of its predictions by comparing against monitoring data of air dose rates in Fukushima Prefecture. The tool is applied for understanding reductions in dose rate seen in North-East Japan in terms of migration of radiocesium within soil, and for evaluating different soil remediation options for contaminated farmlands.

\section{\label{sec:Methods}Methods}

\subsection{\label{ssec:Tool}Tool to evaluate air dose rates}

The tool calculates ambient dose equivalent rates at \SI{1}{\metre} above the ground, $\dot{H}^{\ast}(10)$~(\si{\micro \sievert \per \hour})~\citep{ICRP74}, hereafter referred to as air dose rates. The tool consists of conversion factors between \textsuperscript{134}Cs and \textsuperscript{137}Cs activity concentrations in different cells and layers of soil, and their contribution to the air dose rate. This method allows the dose rate to be calculated for any radiocesium depth profile within soil and horizontal distribution of the activity, to the limit of the precision of the discretization of the ground into the different soil volumes.

\begin{figure*}
\centering
\includegraphics[width=0.95\textwidth]{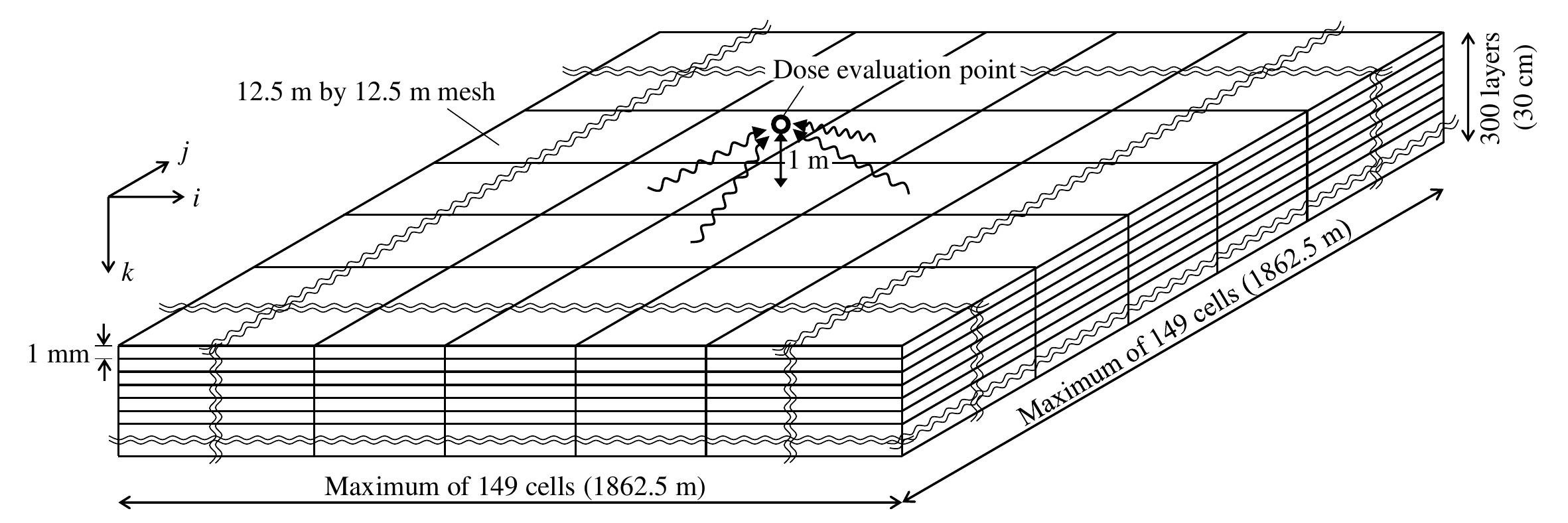}
\caption{\label{fig:geometry}The geometry of the simulations, showing the discretization of soil into small volumes with variable activity concentrations.}
\end{figure*}

The geometry considered is the infinite half-space~\citep{ICRU53} and the land surface is divided into cells by a \num{12.5} by \SI{12.5}{\metre} mesh (Fig.~\ref{fig:geometry}). The tool supports up to \num{149} by \num{149} cells horizontally, which equates to an \num{1862.5} by \SI{1862.5}{\metre} area of land. Up to \num{300} soil layers, each \SI{1}{\milli \metre} thick, are modelled below each cell on the mesh. Thus the maximum depth of radiocesium contamination is \SI{30}{\centi \metre}.

The half-space geometry is a model for open and uniformly flat land. Therefore any natural or man-made geographical features that could significantly alter the air dose rate, such as buildings, hilly topography or dense forests, currently cannot be modelled accurately with the tool. The model also does not consider the effects of ground roughness on air dose rates. These effects are most significant when modelling planes of radionuclides on the surface of the ground in the half-space geometry, such as in the period immediately after fallout deposition. However, they are negligible when modelling radionuclides dispersed within the ground after this initial weathering period has completed~\citep{Jacob1994}, as is the case in this paper. 

The input data for an air dose rate calculation consist of \textsuperscript{134}Cs and \textsuperscript{137}Cs activity concentrations within each discrete volume of soil. The calculation for the air dose rate performed by the tool is given by
\begin{equation}
	\dot{H}^{\ast}(10)=\sum_{n,i,j,k}{A_{\mathrm{v},n,i,j,k}\,c_{n,i,j,k}}\,,
	\label{eq:H10}
\end{equation}
\noindent where $A_{\mathrm{v},n,i,j,k}$~(\si{\becquerel \per \cubic \metre}) is the activity concentration of radiocesium in the soil volume, and $c_{n,i,j,k}$ (\si{\micro \sievert \per \hour} per \si{\becquerel \per \cubic \metre}) is the activity to dose conversion factor for that soil volume. The index $n$ denotes the \textsuperscript{134}Cs or \textsuperscript{137}Cs isotope, indices $i,j$ denote the cell position of the soil volume on the mesh, and $k$ indexes the depth of the volume for layers numbered down from the surface. As the calculation in Eq.~\ref{eq:H10} is a simple sum over all the soil volumes in the problem, the run-time of the tool on a standard desktop computer is about \SI{10}{\second}.

The conversion factors for all the soil volumes were calculated using the Particle and Heavy Ion Transport code System (PHITS) (version \num{2.64} - \citet{Sato2013}). PHITS is a Monte Carlo radiation transport code. The conversion factors represent the dose rate at \SI{1}{\metre} above the middle of the central cell on the mesh per unit activity concentration within that volume of soil. The density of soil was $\rho_\mathrm{s}=1.6$~\si{\gram \per \cubic \centi \metre} and air was $\rho_\mathrm{a}=0.0012$~\si{\gram \per \cubic \centi \metre}. The soil and air chemical compositions followed~\citet{Eckerman1993}. The \textsuperscript{134}Cs and \textsuperscript{137}Cs emission spectra were drawn from~\citet{NuDat2}. Note that the \textsuperscript{137}Cs energy lines in \citet{NuDat2} include the contribution from the short-lived daughter product \textsuperscript{137m}Ba. In each case the source region was scaled to a vertical line and the detectors transformed to planes to maximize the computational efficiency of the Monte Carlo simulations~\citep{Namito2012}.

\subsection{\label{ssec:Transforming}Transforming measured activity depth profiles for input into the tool}

The tool cannot accurately simulate scenarios where there is significant variation in the soil density horizontally across the simulation region, as a constant soil density ($\rho_\mathrm{s}=1.6$~\si{\gram \per \cubic \centi \metre}) was employed in the PHITS simulations. However, other constant soil densities ($\rho_\mathrm{s}\neq 1.6$~\si{\gram \per \cubic \centi \metre}) or soils with varying density as a function of depth (i.e.~$\rho_\mathrm{s}(z)$, where $z$~(\si{\centi \metre}) is the depth in soil from the ground surface) can be simulated. The solution is to transform the depth coordinate of the source activity depth profiles using the mass depth~\citep{ICRU53}. The mass depth, $\zeta$~(\si{\gram \per \square \centi \metre}), is defined as
\begin{equation}
	\zeta(z)=\int_0^{z}{\rho_\mathrm{s}(z')\,\mathrm{d}z'}\,.
	\label{eq:zeta}
\end{equation}

A contamination depth profile per unit soil mass measured in a field survey, $A_\mathrm{m}(\tilde{z})$~(\si{\becquerel \per \kilo \gram}), can be recast using Eq.~\ref{eq:zeta} into a function of soil mass depth
\begin{equation}
	A_\mathrm{m}(\zeta)=A_\mathrm{m}\left(\int_0^{\tilde{z}}{\rho_\mathrm{s}(\tilde{z}')\,\mathrm{d}\tilde{z}'}\right)\,,
	\label{eq:A_m_zeta}
\end{equation}
\noindent where $\rho_\mathrm{s}(\tilde{z})$ is the density-depth profile of the soil measured from field samples. Here tildes are used to distinguish the depth coordinate in the field, $\tilde{z}$~(\si{\centi \metre}), from the depth coordinate $z$ applicable for inputting data into the tool.

The activity profile as a function of mass depth, $A_\mathrm{m}(\zeta)$, can be transformed into a function of $z$ applicable for the tool's constant soil density conversion factors, by reapplying Eq.~\ref{eq:zeta}:
\begin{equation}
	A_\mathrm{m}(z)=A_\mathrm{m}(\zeta/1.6)\,.
	\label{eq:A_m_z}
\end{equation}

Finally, an activity concentration depth profile ($A_\mathrm{v}(z)$) for inputting into the tool can be obtained by multiplying $A_\mathrm{m}(z)$ by the constant soil density
\begin{equation}
	A_\mathrm{v}(z)=1000 \cdot 1.6 \cdot A_\mathrm{m}(z)\,.
	\label{eq:A_v_z}
\end{equation}
\noindent The factor of \num{1000} ensures $A_\mathrm{v}(z)$ has units of \si{\becquerel \per \cubic \metre}.

\subsection{\label{ssec:Problems}Scenarios considered}

\subsubsection{\label{sssec:undisturbed}Dose rates above flat, undisturbed fields}

Field survey teams have measured depth profiles of radioactive cesium in soil at approximately \num{80} locations near to FDNPP since December 2011~\citep{Matsuda2015}. The samples were taken from sites at wide, flat areas of land and at least \SI{5}{\metre} from buildings and trees. Soil samples were collected using a scraper plate. This apparatus was used to remove individual soil layers with thickness between \SIrange{0.5}{3}{\centi \metre} and with increasing depth from the ground surface for radiochemical analysis. Properties analyzed included the in situ density ($\rho_\mathrm{s}(\tilde{z})$) and the \textsuperscript{134}Cs and \textsuperscript{137}Cs activity per unit mass ($A_\mathrm{m}(\tilde{z})$) of each soil layer. All survey data are published online~\citep{JAEA2015a}.

\begin{table*}
\caption{\label{tab:survey_periods}Details of the soil sampling campaigns and sites used for air dose rate predictions.}
\centering
\begin{tabular}{ l l l l }\hline
\begin{tabular}[t]{@{}l@{}}Soil sampling\\campaign\end{tabular} & \begin{tabular}[t]{@{}l@{}}Dates\\\end{tabular} &	\begin{tabular}[t]{@{}l@{}}Number of\\sites used\end{tabular}	& \begin{tabular}[t]{@{}l@{}}Number of sites with\\hyperbolic secant fits\end{tabular} \\ \hline
1\textsuperscript{st} &	\begin{tabular}[t]{@{}l@{}}Dec \numrange{12}{22}, 2011\\and Apr \numrange{17}{19}, 2012\end{tabular}	&	41 &	8 	\\
2\textsuperscript{nd} &	Aug 21 to Sep 26, 2012 						&	82 &	13 	\\
3\textsuperscript{rd} &	Nov 26 to Dec 26, 2012 						&	81 &	28 	\\
4\textsuperscript{th} &	Jun \numrange{4}{27}, 2013 										&	80 &	12 	\\
5\textsuperscript{th} &	Oct 28 to Nov 29, 2013 						&	79 &	23 	\\ \hline
\end{tabular}
\end{table*}

\begin{figure}
\centering
\includegraphics[width=0.4\textwidth]{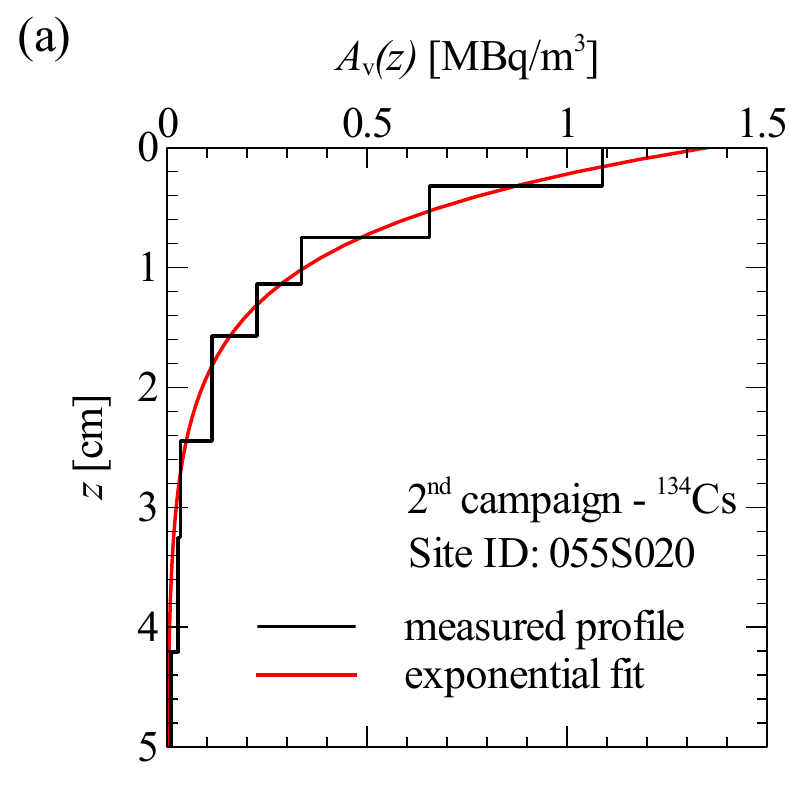} \includegraphics[width=0.4\textwidth]{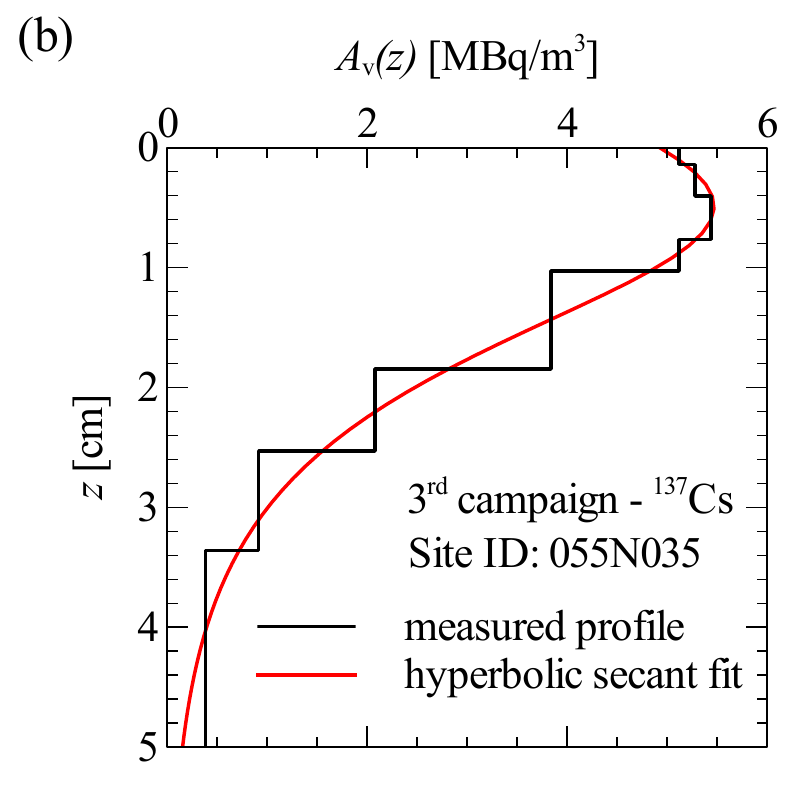}
\caption{\label{fig:input_profiles}Examples of depth distributions of \textsuperscript{134}Cs and \textsuperscript{137}Cs inputted to the tool for dose rate calculations. (a) Exponential depth profile, and (b) hyperbolic secant depth profile. Site identification codes (IDs) follow \citet{Matsuda2015}.}
\end{figure}

We used our calculation tool to predict the air dose rate at each sampling site based on the soil activity measurements. We then compared the results with \SI{1}{\metre} air dose rates measured in the field using hand-held survey meters. Data over five soil sampling campaigns were considered in the analysis. The dates of the campaigns are listed in Table~\ref{tab:survey_periods}. The measured activity depth profiles ($A_\mathrm{m}(\tilde{z})$) were scaled into activity concentration profiles applicable for the tool ($A_\mathrm{v}(z)$) using the procedure described in section~\ref{ssec:Transforming}. Examples of the processed depth profiles for two sites are shown in Fig.~\ref{fig:input_profiles} (black lines).

As a scraper plate sample was taken at only one point on the ground per location visited in each soil sampling campaign, it was assumed that the measured soil activity depth distribution applied homogeneously across the whole region simulated by the tool. Explicitly, the activity concentrations for each soil layer inputted to the tool were identical across all cells on the simulation mesh.

We also considered a second method for evaluating air dose rates from the soil activity samples, based on modelling empirical fits to the activity depth profiles. \citet{Matsuda2015} characterized the activity depth profiles as a function of mass depth ($A_\mathrm{m}(\zeta)$) by fitting exponential and hyperbolic secant functions. The exponential depth distribution is
\begin{equation}
	A_\mathrm{m}(\zeta)=A_{\mathrm{m},0}\exp{(-\zeta/\beta)}\,,
	\label{eq:exp}
\end{equation}
\noindent where $A_{\mathrm{m},0}$~(\si{\becquerel \per \kilo \gram}) is the activity per unit soil mass at the ground surface and $\beta$~(\si{\gram \per \square \centi \metre}) is the relaxation mass depth that characterizes the degree of fallout penetration into the soil. Figure \ref{fig:input_profiles}(a) shows a fit of the exponential function to soil layer activity measurements at one site. The total inventory of contamination per unit area of land for this distribution is:
\begin{equation}
	A_\mathrm{inv}= 10\beta A_{\mathrm{m},0}\,.
	\label{eq:inv_exp}
\end{equation}
\noindent The factor of 10 ensures $A_\mathrm{inv}$ has units of \si{\becquerel \per \square \metre}. The exponential distribution is a satisfactory model for the soil activity depth profile for the first few years after fallout deposition~\citep{ICRU53}.

\citet{Matsuda2015} observed that some of the measured depth profiles display a maximum in the radiocesium concentration below the ground surface. They proposed fitting a hyperbolic secant function to these depth profiles, as this function can reproduce a peak in activity concentration below the surface. The hyperbolic secant function is
\begin{equation}
	A_\mathrm{m}(\zeta)=A_{\mathrm{m},0}\cosh{(\zeta_0/\beta)}\mathrm{sech}(-(\zeta-\zeta_0)/\beta)\,.
	\label{eq:sech}
\end{equation}
\noindent Again $A_{\mathrm{m},0}$~(\si{\becquerel \per \kilo \gram}) is the activity per unit soil mass at the ground surface, and $\beta$~(\si{\gram \per \square \centi \metre}) is a parameter characterizing the length scale of the distribution. The peak in activity occurs at the mass depth $\zeta_0$ (\si{\gram \per \square \centi \metre}) below the surface. The hyperbolic secant function converges to an exponential distribution at large mass depths. Figure \ref{fig:input_profiles}(b) shows a fit of the hyperbolic secant function to a measured depth distribution. The total radionuclide inventory per unit area for the hyperbolic secant distribution is
\begin{eqnarray}
	A_\mathrm{inv}&=&20\beta A_{\mathrm{m},0} \cosh⁡{(\zeta_0/\beta)} [(\pi/4)\nonumber\\
	&&+\tan^{-1}⁡{(\tanh{(\zeta_0/(2\beta))})}]\,.
		\label{eq:inv_sech}
\end{eqnarray}

We followed \citet{Matsuda2015} and fitted the exponential and hyperbolic secant distributions to measured depth profiles. The hyperbolic secant function was used for profiles displaying a peak in activity below the surface, and the exponential function otherwise. Table~\ref{tab:survey_periods} lists the number of sites in this analysis and the number of fits with the hyperbolic secant function. Examples of the fitted distributions are shown for two sites in Fig.~\ref{fig:input_profiles} (red lines)

We discounted from the analysis any depth profiles showing signs of soil mixing or disturbance~\citep{Matsuda2015}. Soil disturbance included land cultivation and decontamination work. We also discounted sites where the air dose rate was not measured with a survey meter at the time of collecting the soil samples.

Modelling both the measured and the empirical fits for the soil depth profiles yielded two predictions for the air dose rate at each site. The air dose rate was calculated as the sum of contributions from \textsuperscript{134}Cs and \textsuperscript{137}Cs, and an additional \SI{0.05}{\micro \sievert \per \hour} contribution representing the background dose rate from natural radionuclides~\citep{Mikami2015b}.

\subsubsection{\label{sssec:Evolution}Evolution of air dose rates}

In addition to the soil sampling campaigns, JAEA and partners have been measuring air dose rates with hand-held survey meters at thousands of locations across Fukushima Prefecture, including locations with flat, undisturbed fields~\citep{Saito2015a, Mikami2015b}. The monitoring results show that dose rates at these locations decreased faster than expected by just the physical decay of \textsuperscript{134}Cs and \textsuperscript{137}Cs~\citep{Saito2015}. \citet{Mikami2015a} demonstrated that, for the period between March 2012 and December 2012, relatively little migration of the \textsuperscript{137}Cs inventory away from these fields occurred. \citet{Mikami2015b} explained the decrease in dose rates, beyond what could be expected by radioactive decay alone, by the downward migration of radioactive cesium into the soil.

The relaxation mass depth, $\beta$, characterizes the penetration of fallout into soil for the exponential distribution (Eq.~\ref{eq:exp}). Conversion coefficients published for various values of $\beta$ can be used to evaluate the \SI{1}{\metre} ambient dose equivalent rate given the radionuclide inventory per unit area of soil~\citep{Saito2014}. In contrast, two parameters characterize the penetration of the radionuclides within soil for the hyperbolic secant distribution - a relaxation mass depth $\beta$ and a mass depth $\zeta_0$ for the peak in activity concentration below the surface (Eq.~\ref{eq:sech}).

To allow direct comparison between exponential and hyperbolic secant depth profiles, \citet{Matsuda2015} proposed an effective relaxation mass depth parameter, $\beta_\mathrm{eff}^{\dot{K}}$~(\si{\gram \per \square \centi \metre}), for the hyperbolic secant distribution. The effective relaxation mass depth is defined as the value $\beta$ of an exponential depth distribution yielding the same air kerma rate at \SI{1}{\metre} as the hyperbolic secant distribution ($\dot{K}$ - \si{\micro \gray \per \hour}), given an identical inventory of fallout radionuclides in both distributions (i.e.~$A_\mathrm{inv}$ is equal for both distributions).

In this study we used our calculation tool to calculate effective relaxation mass depths for the hyperbolic secant fits over the five soil sampling campaigns. The effective relaxation mass depths were calculated by matching $\dot{H}^{\ast}(10)$ from each hyperbolic secant distribution to an exponential distribution of equal inventory, i.e.~$\beta_\mathrm{eff}^{\dot{H}^{\ast}(10)}$~(\si{\gram \per \square \centi \metre}). Note that the definition of effective relaxation mass depth means that $\beta_\mathrm{eff}^{\dot{H}^{\ast}(10)}$ for an exponential distribution is equal to the relaxation mass depth ($\beta$) of the distribution.

\begin{table*}
\caption{\label{tab:beta_eff}Data for $\beta_\mathrm{eff}^{\dot{H}^{\ast}(10)}$ over the five soil sampling campaigns.}
\centering
\begin{tabular}{ l l l l l l l }\hline
\begin{tabular}[t]{@{}l@{}}Soil sampling\\campaign\end{tabular} & \begin{tabular}[t]{@{}l@{}}Number of\\sites used\end{tabular} & 
\begin{tabular}[t]{@{}l@{}}Number of sites with\\hyperbolic secant fits\end{tabular} & 
\begin{tabular}[t]{@{}l@{}}Mean $\beta_\mathrm{eff}^{\dot{H}^{\ast}(10)}$\\(\si{\gram \per \square \centi \metre})\end{tabular} &
\begin{tabular}[t]{@{}l@{}}Median $\beta_\mathrm{eff}^{\dot{H}^{\ast}(10)}$\\(\si{\gram \per \square \centi \metre})\end{tabular} & 
\begin{tabular}[t]{@{}l@{}}Min $\beta_\mathrm{eff}^{\dot{H}^{\ast}(10)}$\\(\si{\gram \per \square \centi \metre})\end{tabular} & 
\begin{tabular}[t]{@{}l@{}}Max $\beta_\mathrm{eff}^{\dot{H}^{\ast}(10)}$\\(\si{\gram \per \square \centi \metre})\end{tabular} \\ \hline
1\textsuperscript{st} &	83 &	12 &	1.13 &	0.93 &	0.24 &	5.95	\\
2\textsuperscript{nd} &	82 &	13 &	1.41 &	1.00 &	0.11 &	8.72	\\
3\textsuperscript{rd} &	81 &	28 &	1.56 &	1.23 &	0.43 &	10.36	\\
4\textsuperscript{th} &	80 &	12 &	1.64 &	1.36 &	0.29 &	7.73	\\
5\textsuperscript{th} &	79 &	23 &	2.17 &	1.85 &	0.38 &	6.41	\\ \hline
\end{tabular}
\end{table*}

We calculated arithmetic mean, median, minimum and maximum $\beta_\mathrm{eff}^{\dot{H}^{\ast}(10)}$ values for each of the five soil sampling campaigns (Table~\ref{tab:beta_eff}). More sites from the first soil sampling campaign could be used in this analysis than for the dose rate predictions (c.f.~Table~\ref{tab:beta_eff} with Table~\ref{tab:survey_periods}), as it was not necessary to have a field survey measurement of the air dose rate in order to calculate $\beta_\mathrm{eff}^{\dot{H}^{\ast}(10)}$.

\begin{table*}
\caption{\label{tab:reduction}Details of the air dose rate surveys and results for the models of dose rate reductions. \textsuperscript{a}Assumed value -- see text for details.}
\centering
\begin{tabular}{ l l@{\extracolsep{10pt}} l@{\extracolsep{8pt}} l@{\extracolsep{10pt}} l@{\extracolsep{8pt}} l@{\extracolsep{10pt}} l@{\extracolsep{8pt}} l }\hline
\multicolumn{2}{l}{Air dose rate survey} & \multicolumn{2}{l}{Soil sampling results} & \multicolumn{2}{l}{Dose rate measurements} & \multicolumn{2}{l}{Models} \\ \cline{1-2}\cline{3-4}\cline{5-6}\cline{7-8}
Campaign & Dates & Campaign & \begin{tabular}[t]{@{}l@{}}Mean $\beta_\mathrm{eff}^{\dot{H}^{\ast}(10)}$\\(\si{\gram \per \square \centi \metre})\end{tabular} & \begin{tabular}[t]{@{}l@{}}$\dot{H}^{\ast}(10)$ - \num{0.05}\\(\si{\micro \sievert \per \hour})\end{tabular} & \begin{tabular}[t]{@{}l@{}}Relative\\change\end{tabular} & \begin{tabular}[t]{@{}l@{}}Decay\\only\end{tabular}   & \begin{tabular}[t]{@{}l@{}}Decay \&\\migration\end{tabular}\\ \hline
1\textsuperscript{st} campaign																& \begin{tabular}[t]{@{}l@{}}Jun 4 to\\Jul 8, 2011\end{tabular}						& - &	1.00\textsuperscript{a} & 1.25	&	1.0		&	1.0			&	1.0		\\
2\textsuperscript{nd} campaign																& \begin{tabular}[t]{@{}l@{}}Dec 13, 2011 to\\May 29, 2012\end{tabular}	& 1\textsuperscript{st} &	1.13 & 1.01	&	0.81	&	0.84		&	0.82	\\
\begin{tabular}[t]{@{}l@{}}1\textsuperscript{st} part of 3\textsuperscript{rd}\\campaign\end{tabular}	&	\begin{tabular}[t]{@{}l@{}}Aug 14 to\\Sep 7, 2012\end{tabular}		&	2\textsuperscript{nd} &	1.41 & 0.84	&	0.67	&	0.76		&	0.70	\\
\begin{tabular}[t]{@{}l@{}}2\textsuperscript{nd} part of 3\textsuperscript{rd}\\campaign\end{tabular}	&	\begin{tabular}[t]{@{}l@{}}Nov 5 to\\Dec 7, 2012\end{tabular}		&	3\textsuperscript{rd} &	1.56 &	0.78	&	0.62	&	0.72		&	0.65	\\
\begin{tabular}[t]{@{}l@{}}1\textsuperscript{st} part of 4\textsuperscript{th}\\campaign\end{tabular}	&	\begin{tabular}[t]{@{}l@{}}Jun 3 to\\Jul 4, 2013\end{tabular}						&	4\textsuperscript{th} &	1.64 &	0.64	&	0.51	&	0.64		&	0.57	\\
\begin{tabular}[t]{@{}l@{}}2\textsuperscript{nd} part of 4\textsuperscript{th}\\campaign\end{tabular}	&	\begin{tabular}[t]{@{}l@{}}Oct 28 to\\Dec 2, 2013\end{tabular}		&	5\textsuperscript{th} &	2.17 &	0.55	&	0.44	&	0.59		&	0.48	\\ \hline
\end{tabular}
\end{table*}

We considered the decrement of the components of the air dose rate attributable to radioactive cesium fallout, i.e.~$\dot{H}^{\ast}(10)$ measurements minus a \SI{0.05}{\micro \sievert \per \hour} contribution from natural background radiation, over the first four air dose rate surveys~\citep{JAEA2015a}. The dates of the air dose rate surveys and the mean air dose rates at flat, undisturbed fields are listed in Table~\ref{tab:reduction}.

We modelled the decrement in dose rates due to radioactive decay and cesium migration deeper within soil. First, we matched the mean $\beta_\mathrm{eff}^{\dot{H}^{\ast}(10)}$ values from the soil sampling campaigns to the periods of the air dose rate surveys (Table~\ref{tab:reduction}). We then modelled exponential distributions with $\beta$ parameters equal to the mean $\beta_\mathrm{eff}^{\dot{H}^{\ast}(10)}$ values with the tool. The inventories supplied were decay corrected to dates at the middle of each air dose rate survey period. The decay corrections assumed an activity ratio of released \textsuperscript{134}Cs and \textsuperscript{137}Cs from FDNPP of \num{1.00} on March 11, 2011~\citep{UNSCEAR2014}. The results show little sensitivity to plausible alternatives (in the range \numrange{0.90}{1.08}) for this initial activity ratio. The calculated dose rates were then normalized to June 21, 2011, the date at the middle of the first air dose rate survey (Table~\ref{tab:reduction}), for comparison with the measured dose rates. As no scraper plate soil samples were available for the period of the first air dose rate survey (June 4 to July 8, 2011), a $\beta_\mathrm{eff}^{\dot{H}^{\ast}(10)}$ value of \SI{1.00}{\gram \per \square \centi \metre} was assumed as applicable for this period~\citep{Mikami2015b}. The sensitivity of the results upon this assumption is checked in the results section.

\subsubsection{\label{sssec:Spatial}Spatial variability in soil activity levels}

\begin{figure*}
\centering
\includegraphics[width=0.8\textwidth]{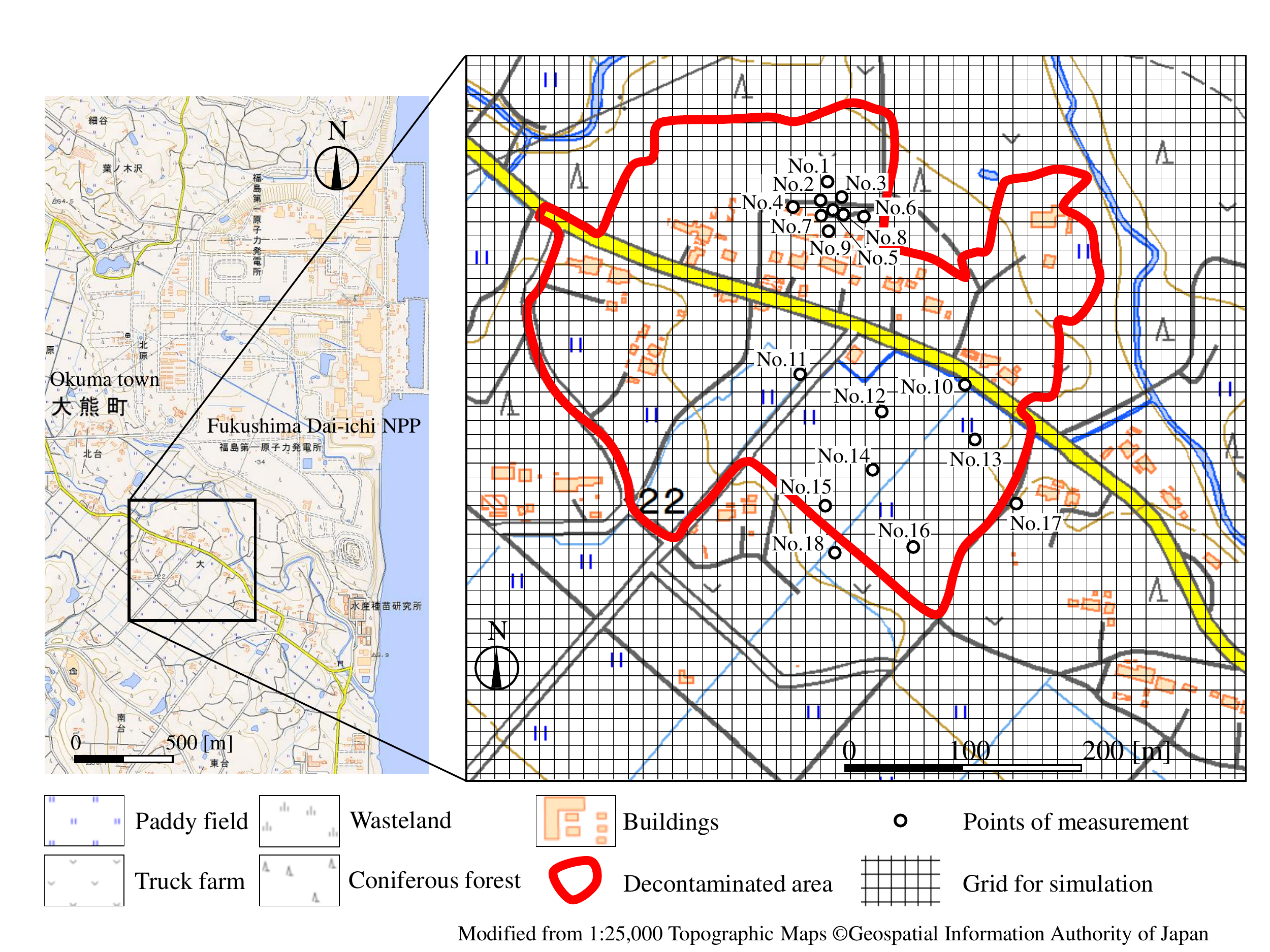}
\caption{\label{fig:ottozawa}The decontamination boundary and soil sampling locations at Ottozawa.}
\end{figure*}

The calculations with the tool in sections~\ref{sssec:undisturbed} and \ref{sssec:Evolution} assumed spatially uniform radiocesium distributions, as only one soil sample was available at each location. We considered the effect of spatial variability in the radiocesium distribution on evaluating dose rates by studying the Ottozawa area. The area lies within \SI{2}{\kilo \metre} of FDNPP and soil samples were taken at multiple locations across the paddy fields and scrubland in the area. Figure \ref{fig:ottozawa} shows a map of the area and the soil sampling locations.

This area was remediated between November 2011 and May 2012 as part of a decontamination pilot project coordinated by JAEA and is now subject to long-term environmental monitoring~\citep{JAEA2015b}. Remediation consisted of removing the top \SI{5}{\centi \metre} of topsoil from paddy fields and areas around residential buildings, and cleaning road and building surfaces. The air dose rates ranged from \SIrange{22}{263}{\micro \sievert \per \hour} before decontamination, and dropped to between \SIrange{4}{110}{\micro \sievert \per \hour} afterwards. Decontamination of this area was studied numerically by \citet{Hashimoto2014}.

The soil samples and air dose rates were taken on July 24, 2014 at \num{18} locations across the area. Soil samples were collected by inserting a cylindrical plastic cup (U-8 type, \SI{58}{\milli \metre} internal height, \SI{50}{\milli \metre} internal diameter) into the topsoil and collecting the soil contents into plastic bags~\citep{Onda2015}. One sample was taken at locations \numrange{1}{16}, while five samples were collected for locations \num{17} and \num{18}. Locations \num{17} and \num{18} lie outside the decontaminated area.

\begin{figure}
\centering
\includegraphics[width=0.4\textwidth]{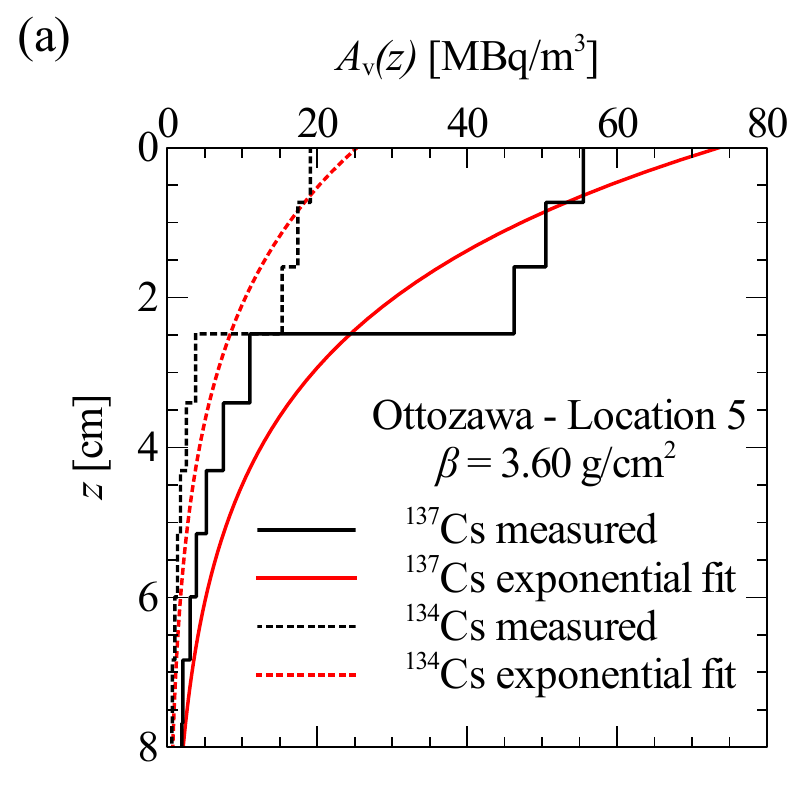} \includegraphics[width=0.4\textwidth]{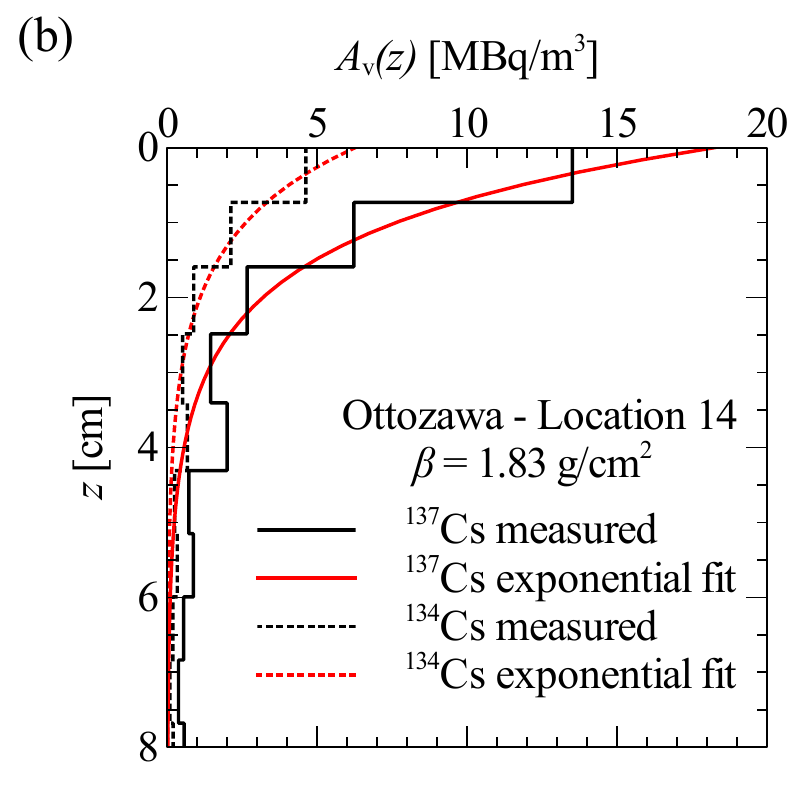}
\caption{\label{fig:ottozawa_profiles}Measured soil activity depth distributions and exponential fits for the Ottozawa area. (a) Location~\num{5}, and (b) location~\num{14}.}
\end{figure}

The \textsuperscript{134}Cs and \textsuperscript{137}Cs depth distributions at locations \num{5} and \num{14} were determined by using a scraper plate to remove \SI{1}{\centi \metre} thick soil layers down to a maximum depth of \SI{10}{\centi \metre}. The activity per unit soil mass in each layer was measured using a high resolution gamma spectrometer. Unfortunately due to oversight we did not measure the in situ densities at the time of collecting the soil samples. Therefore we had to make an assumption for the layer densities. We chose densities equal to the mean densities of the soil layers collected over the five scraper plate sampling campaigns described in section~\ref{sssec:undisturbed}. The measured activity depth profiles at Ottozawa were exponential to a reasonable approximation (Fig.~\ref{fig:ottozawa_profiles}).

Scraper plate analyses of the activity depth distributions were not performed at the other locations (locations \numrange{1}{18}, excluding \num{5} and \num{14}). An exponential depth distribution was assigned to these locations based on the $\beta$ value applicable at the nearest of locations \num{5} and \num{14} to the site. Locations \numrange{1}{9} were thus assigned an exponential depth distribution with $\beta=3.60$~\si{\gram \per \square \centi \metre}, and locations \numrange{10}{18} an exponential depth distribution with $\beta=1.83$~\si{\gram \per \square \centi \metre}. The total inventory per unit area, $A_\mathrm{inv}$, was inferred by correcting the cylindrical cup activity measurement for radioactivity at depths greater than \SI{58}{\milli \metre} as given by the exponential distribution. The inventory for locations \numlist{5;14;17;18}, where multiple soil samples were taken, was taken to be the mean over the various samples.

Two strategies were used to predict the air dose rate at locations \numrange{1}{18}. The first strategy assumed that the radiocesium distribution was spatially homogeneous. The inventory and depth distribution for that location was applied uniformly across the simulation region.

\begin{figure*}
\centering
\includegraphics[width=0.7\textwidth]{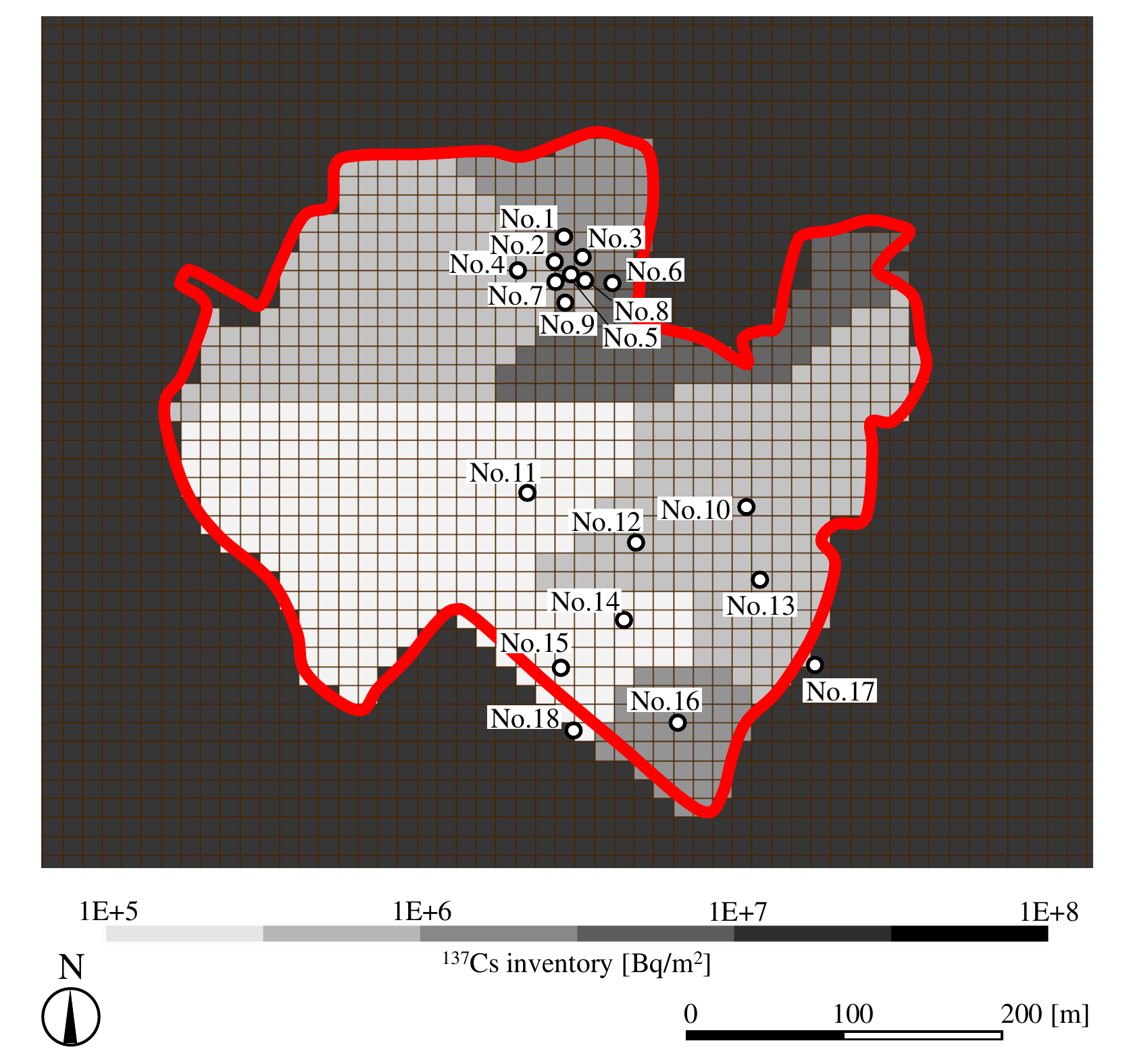}
\caption{\label{fig:ottozawa_inventories}\textsuperscript{137}Cs inventories assigned to cells across the Ottozawa area for simulation of air dose rates at locations \numrange{1}{18}.}
\end{figure*}

The second strategy was to model the spatial heterogeneity in soil activity levels, as revealed by the soil samples at the other locations. A \num{12.5} by \SI{12.5}{\metre} mesh was overlaid onto a map of the area (Fig.~\ref{fig:ottozawa}). Cells containing a soil sampling location were assigned the inventory and relaxation mass depth for that sample. We adopted a simple interpolation method to assign inventories and relaxation mass depths to the other cells on the mesh. The inventories and $\beta$ values were set equal to the values applicable at the nearest cell hosting a sampling location. Cells equidistant from more than one sampling location were assigned inventories randomly from one of the equidistant locations. Because of a large disparity between soil activity levels inside and outside the bounds of the remediated area, locations outside the remediated area were assigned the inventory of the closest of either location \num{17} or \num{18}. It would also be possible to employ other interpolation techniques to assign inventories to cells without soil samples, for example based on inverse distance weighting techniques or Kriging~\citep{TECDOC1363}.

The assigned inventories for all cells across the area are depicted in Fig.~\ref{fig:ottozawa_inventories}. The mesh size simulated in the tool was \num{149} by \num{149} cells for both dose rate prediction methods.

\subsubsection{\label{sssec:decon}Evaluation of farmland soil remediation methods}

\begin{figure*}
\centering
\includegraphics[width=0.35\textwidth]{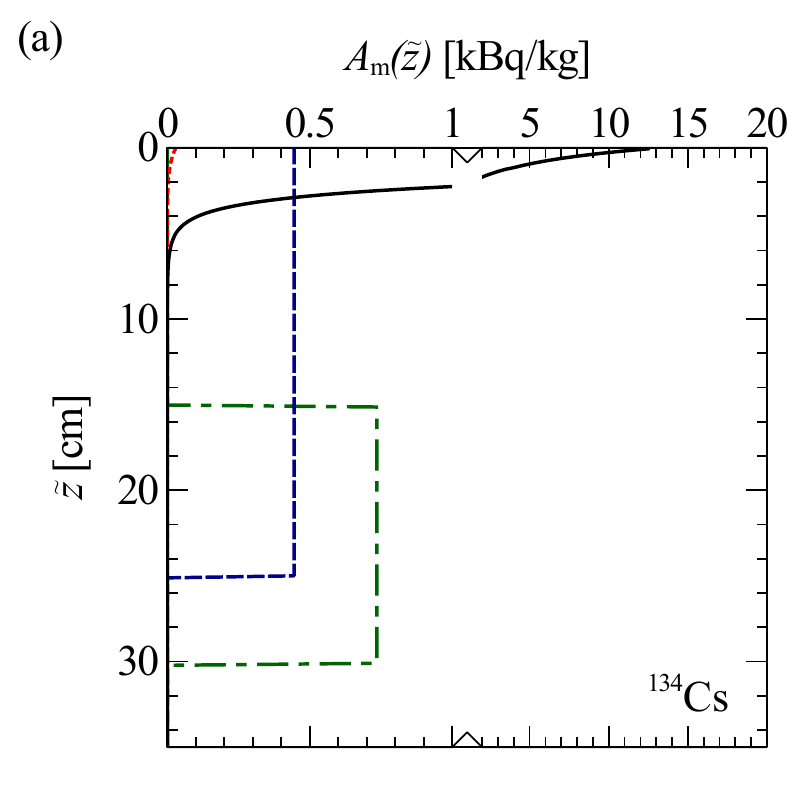} \includegraphics[width=0.55125\textwidth]{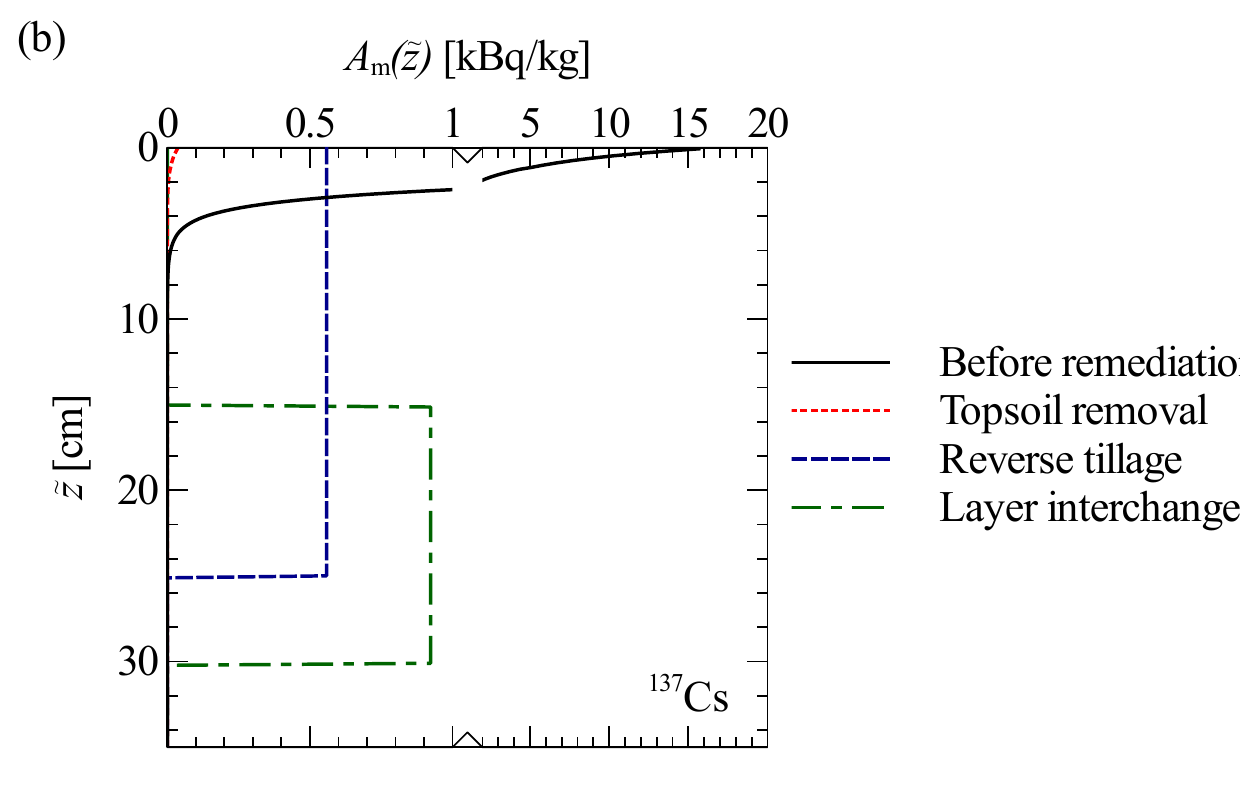}
\caption{\label{fig:decon_profiles}Activity depth distributions of \textsuperscript{134}Cs and \textsuperscript{137}Cs for three farmland soil remediation methods, shown for the in situ depth coordinate $\tilde{z}$. Note break in horizontal axes at \SI{1}{\kilo \becquerel \per \kilo \gram} to show full distribution of activity with depth.}
\end{figure*}

To evaluate different methods for remediating farmland soils, we used the tool to calculate air dose rates after remediation by topsoil removal, reverse tillage, or topsoil-subsoil layer interchange. Figure \ref{fig:decon_profiles} shows a typical exponential depth distribution for \textsuperscript{134}Cs and \textsuperscript{137}Cs within undisturbed farmland soil in Fukushima Prefecture (solid black lines). The \textsuperscript{134}Cs to \textsuperscript{137}Cs activity ratio is applicable on December 01, 2011. This date falls within a pilot project on decontamination techniques, and allows comparison of dose rate predictions from the tool against environmental measurements from the decontamination project~\citep{JAEA2015b}.

The relaxation mass depth of the exponential profile in Fig.~\ref{fig:decon_profiles} is $\beta=1.13$~\si{\gram \per \square \centi \metre}. This follows the result from the first soil depth distribution sampling campaign (Table~\ref{tab:beta_eff}). The air dose rate under these \textsuperscript{134}Cs and \textsuperscript{137}Cs inventories and depth profiles is \SI{1.25}{\micro \sievert \per \hour} before remediation, including a \SI{0.05}{\micro \sievert \per \hour} contribution from natural background radiation.

The different remediation methods alter the activity depth distributions of the farmland soil. Figure \ref{fig:decon_profiles} shows idealized activity depth distributions after topsoil removal, reverse tillage, or topsoil-subsoil layer interchange. We used the tool to evaluate the air dose rate after completion of each of these remediation options.

Topsoil removal involves mechanically stripping the top \SI{5}{\centi \metre} of the soil, and disposing the excavated soil as radioactive waste. The activity profile for the remaining soil is, to a first approximation, the exponential distribution for depths greater than \SI{5}{\centi \metre} prior to decontamination (dotted red lines in Fig.~\ref{fig:decon_profiles}).

Reverse tillage employs a tractor pulled plough to invert the topsoil. The ploughing creates small ridges and furrows on the land surface, which flatten off as the soil weathers and relaxes. We approximated the soil as being homogeneously mixed after this process. Ploughing down to a depth of \SI{25}{\centi \metre} thus results in a constant radioactivity profile initially with depth, followed by the exponential distribution at depths below \SI{25}{\centi \metre} (dashed blue lines in Fig.~\ref{fig:decon_profiles}).

In topsoil-subsoil layer interchange a layer of topsoil is switched with a layer of subsoil. Typically a topsoil layer down to \SI{15}{\centi \metre} is excavated with a digger and this soil is placed aside on a plastic sheet. The next \SI{15}{\centi \metre} of subsoil is then excavated and stored temporarily on adjacent ground. The pit that has been created is refilled by first adding a \SI{15}{\centi \metre} layer of the original topsoil, and then levelling to the ground surface with the excavated subsoil layer. This strategy can be approximated as creating two homogenized layers of activity concentration below the ground surface. The top layer, down to \SI{15}{\centi \metre} depth, contains the activity originally between the depths of \SI{15}{\centi \metre} and \SI{30}{\centi \metre}. The subsequent \SI{15}{\centi \metre} thick layer below contains the activity that was originally in the top \SI{15}{\centi \metre} of soil (green dash-dot lines in Fig.~\ref{fig:decon_profiles}).

\begin{figure}
\centering
\includegraphics[width=0.4\textwidth]{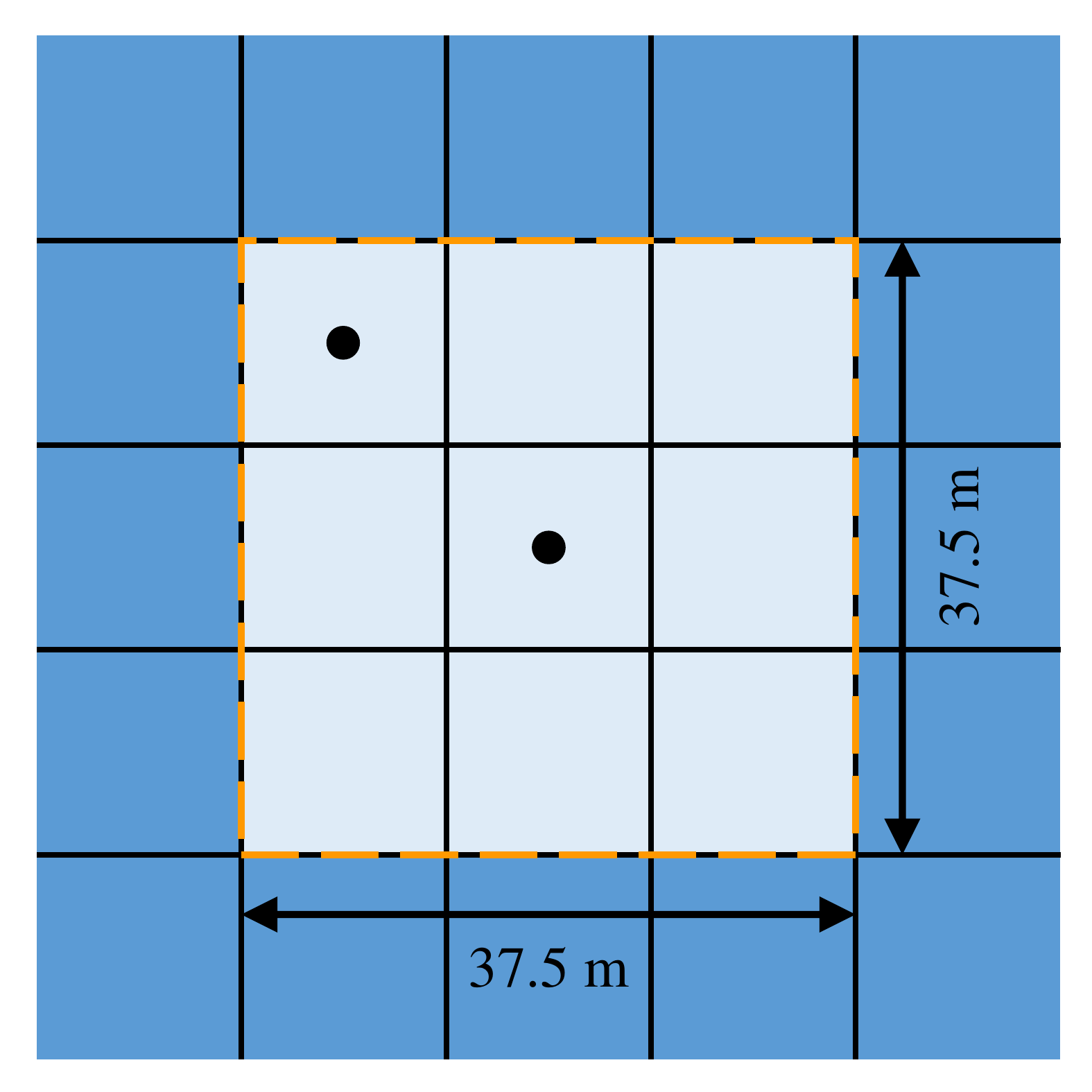}
\caption{\label{fig:remediation_square}Setup of farmland soil remediation simulations: light blue area within orange dashed line is remediated land. Land in the dark blue area outside is not remediated. Dose rates before and after remediation were calculated for the locations marked by black spots.}
\end{figure}

To model these remediation scenarios, we considered remediation of a \num{37.5} by \SI{37.5}{\metre} (\SI{1.4}{\square \kilo \metre}) area of land, equivalent to a \num{3} by \num{3} square of cells on the simulation mesh (Fig.~\ref{fig:remediation_square}). The simulation models consisted of remediated depth distributions within these cells, while the depth distributions outside the area remained unchanged. Reductions in the dose rates were calculated for the center and near to the corners of the remediated square of land. All dose rate evaluations included a \SI{0.05}{\micro \sievert \per \hour} contribution from natural background radiation.

\section{\label{sec:results}Results and discussion}

\subsection{\label{ssec:undisturbed2}Dose rates above flat, undisturbed fields}

\begin{figure*}
\centering
\includegraphics[width=0.4\textwidth]{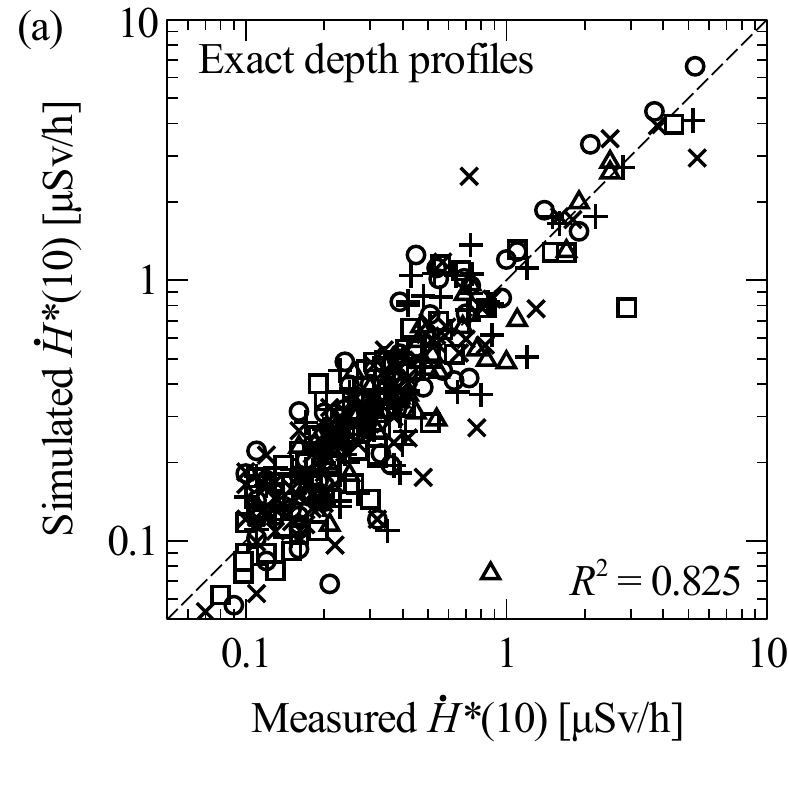} \includegraphics[width=0.575\textwidth]{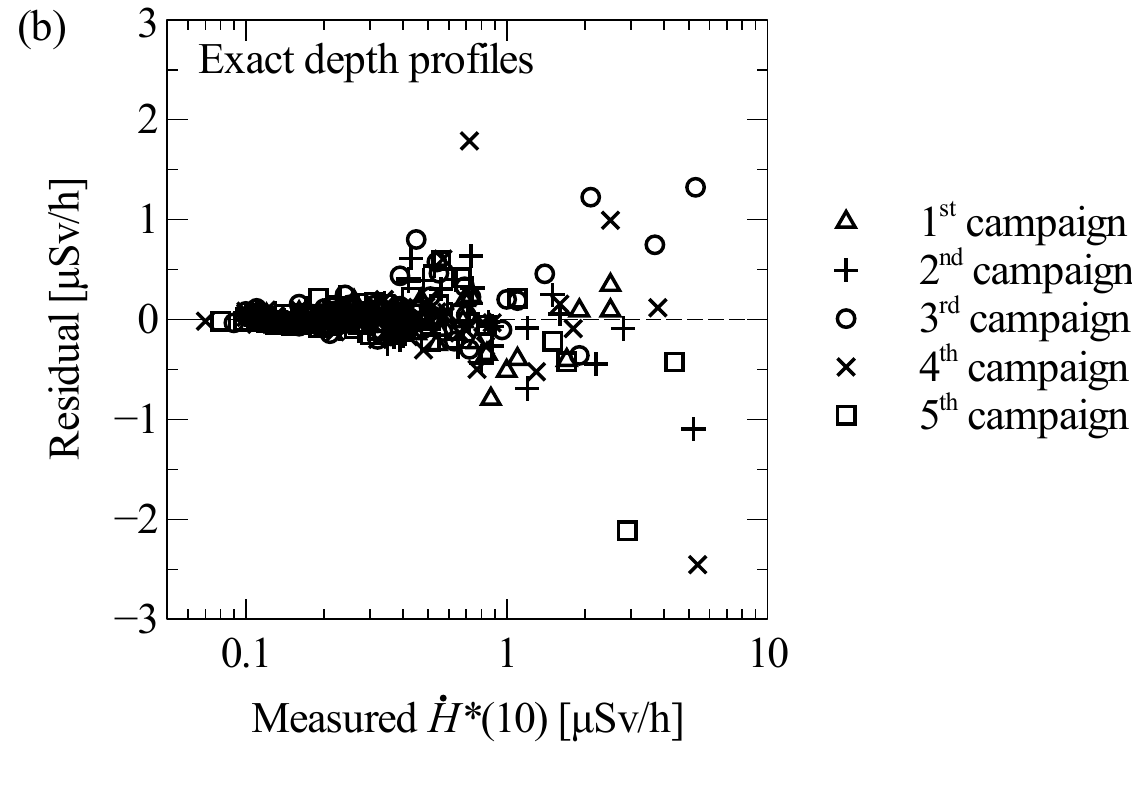}
\caption{\label{fig:scatter_exact}Correlation between measured air dose rates from the soil sampling campaigns and simulation predictions using measured soil depth profiles as inputs. (a) Measurement-prediction correlation. The dotted line indicates $y=x$. (b) Scatter plot of residual errors in the predictions. The dotted line is $y=0$.}
\end{figure*}

The predictions for air dose rates above flat, undisturbed fields made using the measured activity depth profiles compare well with the dose rates measured at the sampling sites, as shown by Fig.~\ref{fig:scatter_exact}(a). The correlation holds over the range of dose rates covered by the dataset (\SIrange{0.09}{5.3}{\micro \sievert \per \hour}). The predicted dose rates are always within a factor of three of the true dose rate, with one exception. At one site a dose rate of \SI{0.87}{\micro \sievert \per \hour} was observed, but the tool predicted \SI{0.075}{\micro \sievert \per \hour}.

The residual differences between the predictions and the measured dose rates are shown in Fig.~\ref{fig:scatter_exact}(b). A positive residual indicates an over-estimation by the tool, and a negative residual, an under-estimate. There is no tendency for the tool to either over-estimate or under-predict dose rates across the range of dose rates measured in the surveys.

\citet{Tyler1996} noted previously that individual soil samples can be poor representations of the mean soil activity across a wide area. The mean free path in air of the primary gamma rays emitted by \textsuperscript{134/137}Cs decay is around \SI{100}{\metre}. \citet{Satoh2014} showed that radioactivity within \SI{500}{\metre} contributes significantly to an air dose rate. Thus, the total volume of soil contributing to the dose rate, down to a depth of \SI{8}{\centi \metre}, is \SI{62800}{\cubic \metre}. As the volume of soil collected down to the same depth with a \SI{15}{\centi \metre} by \SI{30}{\centi \metre} scraper plate is \SI{0.0036}{\cubic \metre}, the sample represents only \num{6e-8} parts of the total soil volume contributing to the air dose rate.

Highly variable \textsuperscript{134}Cs and \textsuperscript{137}Cs activity concentrations are often found between different soil samples taken at the same location. \citet{Saito2015a} confirmed this was the case for soil samples taken in Fukushima Prefecture. The variations are caused by heterogeneity in the fallout deposition, and by scrubbing and concentration of fallout nuclides by local earth surface processes. Therefore, a large sampling uncertainty for the inventory of the total soil volume contributing to the air dose rate should be expected if only a single soil sample is available. We ascribe the sampling uncertainty from the scraper plate measurement as the main source of error in the predictions for the air dose rate shown in Fig.~\ref{fig:scatter_exact}(a).

\begin{figure*}
\centering
\includegraphics[width=0.4\textwidth]{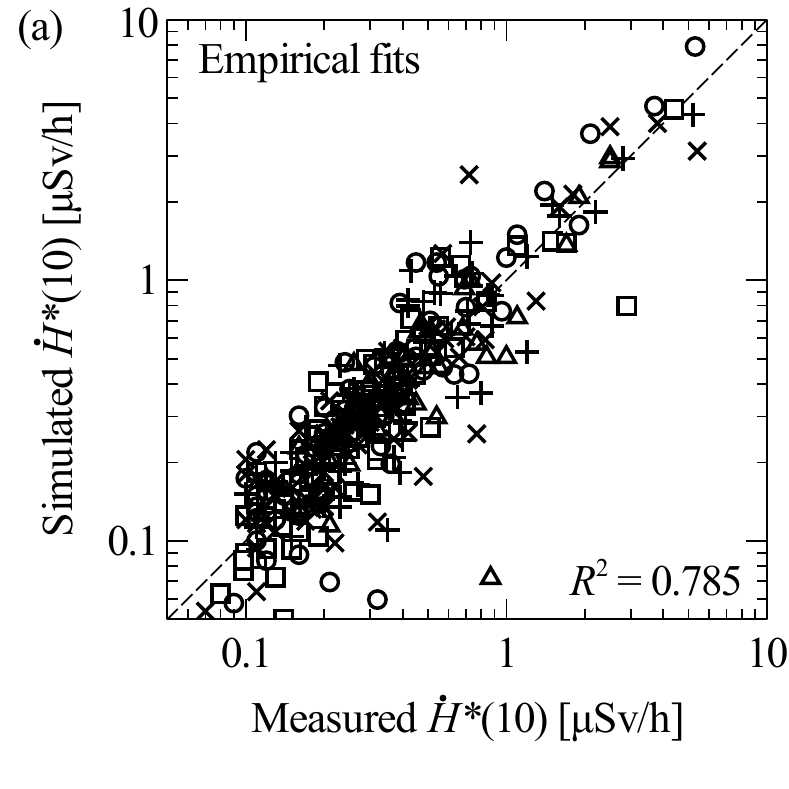} \includegraphics[width=0.575\textwidth]{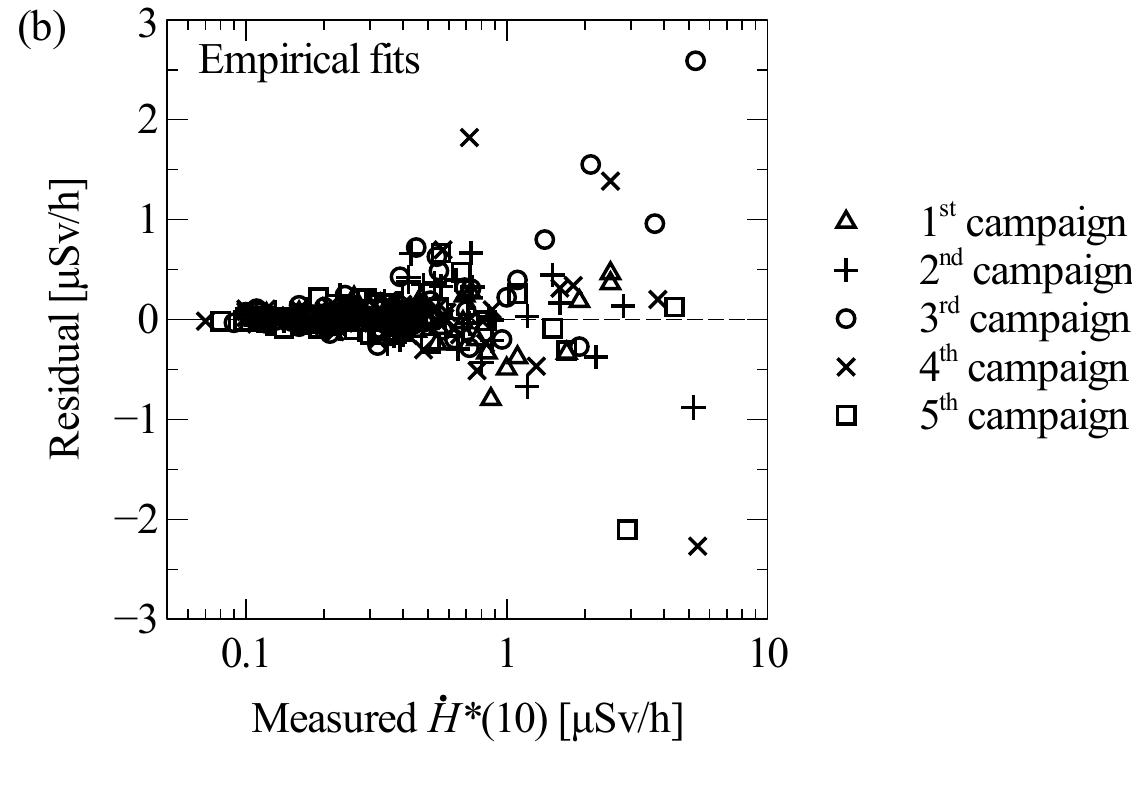}
\caption{\label{fig:scatter_fit}As per Fig.~\ref{fig:scatter_exact}, except showing dose rate predictions made the exponential and hyperbolic secant fits to the measured soil activity depth distributions.}
\end{figure*}

We next considered the quality of the dose rate predictions obtained by modelling the empirical fits to the measured depth profiles (Fig.~\ref{fig:scatter_fit}(a)). The coefficient of determination obtained in this case is slightly lower than the models employing the measured depth profiles directly ($R^2=0.785$ versus \num{0.825}). The slight difference in $R^2$ values is caused by the predictions for the high dose rate locations being slightly less accurate from the models employing the empirical fitting functions. The residual errors for the predictions at these high dose rate locations dominate the squared residuals sum in the calculation of $R^2$, and hence the resulting $R^2$ value.

The residuals for the predictions obtained by modelling the empirical fits are shown in Fig.~\ref{fig:scatter_fit}(b). Excluding the high dose rate locations, the amount of scatter in the residuals is comparable to Fig.~\ref{fig:scatter_fit}(a). Another way to quantify the accuracy of the predictions is to consider the mean absolute percentage error. This statistic is less susceptible to being skewed by the squared residuals for the predictions at the high dose rate locations than $R^2$. The mean absolute percentage error of the predictions made using the exact depth profiles is \SI{29}{\percent}. This compares with a mean absolute percentage error of \SI{30}{\percent} for the predictions obtained by modelling the fitted activity depth profile functions.

The results thus indicate that no significant error is introduced by modelling the empirical fits to the activity depth profiles instead of the measured step-wise profiles. This conclusion necessarily depends on the details of the soil sampling procedure. \citet{Matsuda2015} measured the activity within \SI{0.5}{\centi \metre} layers of topsoil, followed by \num{1} and \SI{3}{\centi \metre} thick layers at deeper depths. If coarser soil layer thicknesses are employed, modelling the empirical fits may yield more accurate predictions than modelling the measured depth profiles, as it is plausible for the empirical fits to offer a better representation of the true activity profile in the soil.

\subsection{\label{ssec:evoluation2}Evolution of air dose rates}

\begin{figure*}
\centering
\includegraphics[width=0.42\textwidth]{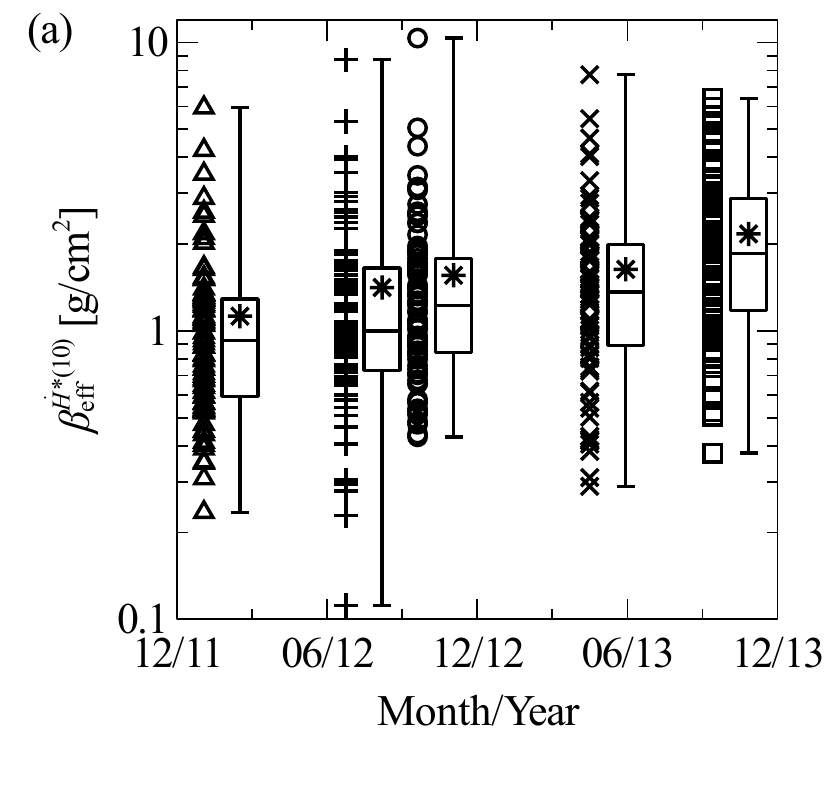} \includegraphics[width=0.445\textwidth]{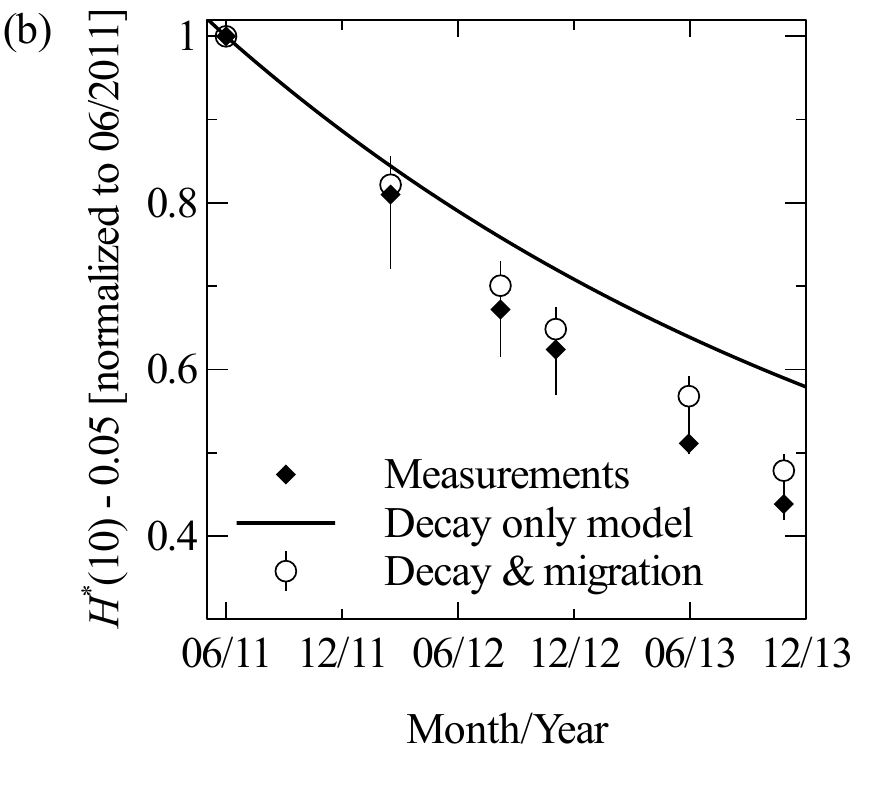}
\caption{\label{fig:evolution}(a) Box and whisker plot showing distribution of $\beta_\mathrm{eff}^{\dot{H}^{\ast}(10)}$ values over the five soil sampling campaigns. The whiskers show the maxima and minima of the distributions. The boxes show the range in between the 25\textsuperscript{th} and 75\textsuperscript{th} percentiles of distributions. The mean values are indicated by asterisks. The full distributions are plotted with symbols, offset to the left of each box and whiskers. (b) Measurements and modelling results for the reduction in air dose rate component attributable to radioactive cesium at locations of flat, undisturbed fields. The measurements (solid diamonds) show the mean air dose rate attributable to radiocesium from the air survey campaigns, normalized to the value at the first air dose rate survey (Table~\ref{tab:reduction}). The vertical bars on the data for the decay and migration model (circles) indicate results when varying $\beta_\mathrm{eff}^{\dot{H}^{\ast}(10)}$ between \SIrange{0.5}{2.0}{\gram \per \square \centi \metre} at the time of the first air dose rate survey (June 2011).}
\end{figure*}

The distributions of $\beta_\mathrm{eff}^{\dot{H}^{\ast}(10)}$ values obtained from the exponential and hyperbolic secant fits to the depth profiles are shown in Fig.~\ref{fig:evolution}(a) for the five soil sampling campaigns. Both the mean and median values of $\beta_\mathrm{eff}^{\dot{H}^{\ast}(10)}$ increase over time (Table~\ref{tab:beta_eff}), indicating that the radiocesium is migrating deeper into the soil.

The component of the mean air dose rate attributable to radiocesium at the flat, undisturbed fields is plotted for the air survey campaigns with solid diamonds in Fig.~\ref{fig:evolution}(b). The data are plotted relative to June 2011, the date of the first air dose rate survey (Table~\ref{tab:reduction}).

The solid line in Fig.~\ref{fig:evolution}(b) represents the decrement in dose rates that would be expected on the basis of radioactive decay of \textsuperscript{134}Cs and \textsuperscript{137}Cs, and without migration of the radiocesium at the sites. The measured dose rates decrease faster than expected by just radioactive decay.

\citet{Mikami2015b} explained the additional reduction in dose rates between June 2011 and December 2012 by migration of the radiocesium fallout deeper into the soil. This trend continued through 2013, as shown by the results of our decay and migration calculations (open circles, Fig.~\ref{fig:evolution}(b)). The decay and migration model results are reasonably consistent with the measurements, although they tend to under-estimate the reduction in dose rates by up to \SI{10}{\percent}.

A source of uncertainty in the decay and migration model is the choice for mean $\beta_\mathrm{eff}^{\dot{H}^{\ast}(10)}$ for the first air dose rate survey campaign (June 4 to July 8, 2011 - Table~\ref{tab:reduction}). The soil sampling campaigns by \citet{Matsuda2015} commenced in December 2011, so cannot provide measurements to derive a mean $\beta_\mathrm{eff}^{\dot{H}^{\ast}(10)}$ value applicable to this period. The circles in Fig.~\ref{fig:evolution}(b) represent the assumption that $\beta_\mathrm{eff}^{\dot{H}^{\ast}(10)}=1.0$~\si{\gram \per \square \centi \metre} in June 2011. \citet{ICRU53} cites $\beta$ values for atmospheric radionuclide fallout in the range \SIrange{0.1}{4}{\gram \per \square \centi \metre} for up to one year after fallout deposition. These results are based on measurements for cesium radioisotopes from Chernobyl fallout in Europe and Western Russia.

\citet{Takahashi2015} measured depth distributions at two grassland sites and three abandoned agricultural fields in Fukushima Prefecture between June \numrange{21}{28}, 2011. They found that the exponential distribution was a good fit for the measured depth profiles, with $\beta$ values in the range \SIrange{0.60}{3.08}{\gram \per \square \centi \metre}. However, they noted that the site giving the highest relaxation mass depth (\SI{3.08}{\gram \per \square \centi \metre}) was pasture land where the soil had been disturbed by cattle grazing. Excluding this site from their dataset yields a mean $\beta$ value from four sites of \SI{1.20}{\gram \per \square \centi \metre}.

To determine the sensitivity of our decay and migration model on the choice for the mean $\beta_\mathrm{eff}^{\dot{H}^{\ast}(10)}$ value for the first air dose rate survey, we considered the effect of varying this parameter in the range \SIrange{0.50}{1.20}{\gram \per \square \centi \metre}. This is a range of values that we consider credible for the period between June 4 and July 8, 2011, based on the previous literature cited and the mean $\beta_\mathrm{eff}^{\dot{H}^{\ast}(10)}$ value of \SI{1.13}{\gram \per \square \centi \metre} derived from the first soil sampling campaign in December 2011 (Table~\ref{tab:beta_eff}). The effect of varying the initial value of  $\beta_\mathrm{eff}^{\dot{H}^{\ast}(10)}$ in this range is shown by vertical bars around circle markers in Fig.~\ref{fig:evolution}(b). The ranges indicated by these bars include the measurements, but do not permit the decay only explanation for the reduction in dose rates. The sensitivity analysis is thus consistent with the conclusion that migration of cesium deeper into the ground was the main cause behind the additional decrement in dose rates.

There are two other factors that could plausibly explain the underestimation of the true dose rate reduction by the model for decay and migration deeper into soil (Fig.~\ref{fig:evolution}(b)). Although \citet{Mikami2015a} suggested that little migration of the radiocesium inventory in the horizontal direction had occurred, within uncertainties their data are consistent with a possible small amount of horizontal migration (on the order of \SIrange{5}{10}{\percent} of the inventory).

Another factor is as follows. Although the sites featuring in the air dose rate surveys were chosen to be flat, open spaces~\citep{Mikami2015b}, certain sites may include urban areas or areas with roads and paved surfaces at the periphery. The wide field of view of environmental radioactivity means that radiocesium within these areas contributes to the air dose rate. As the radiocesium within these areas has a shorter ecological half-life than areas of glassland or agricultural areas~\citep{Kinase2014}, i.e.~the radiocesium is more easily washed away, this could contribute to the underestimation of the dose rate reduction by the models in Fig.~\ref{fig:evolution}(b).

\subsection{\label{ssec:spatial2}Effect of spatial variability in soil activity levels}

\begin{table*}
\caption{\label{tab:ottozawa}Results of soil sampling and air dose rate predictions for Ottozawa area on July 24, 2014. \textbf{Bold} indicates soil samples taken with scraper plate apparatus. Other samples collected with U-8 cup. \textit{Italic} indicates values of $\beta$ inferred from depth distributions at locations \num{5} or \num{14}.}
\centering
\begin{tabular}{ l l l l l l l }\hline
Location & \multicolumn{2}{l}{\begin{tabular}[t]{@{}l@{}}Inventory\\(\si{\mega \becquerel \per \square \metre})\end{tabular}} & \begin{tabular}[t]{@{}l@{}}$\beta$\\(\si{\gram \per \square \centi \metre})\end{tabular} & \begin{tabular}[t]{@{}l@{}}Measured $\dot{H}^{\ast}(10)$\\(\si{\micro \sievert \per \hour})\end{tabular} & \multicolumn{2}{l}{\begin{tabular}[t]{@{}l@{}}Prediction based on assumption\\for Cs distribution (\si{\micro \sievert \per \hour})\end{tabular}} \\ \cline{2-3}\cline{6-7}
& \textsuperscript{134}Cs & \textsuperscript{137}Cs &  &  & Homogeneous & Heterogeneous\\ \hline
1		& 0.643		& 1.89	& \textit{3.60}	& 14.6	& 5.3		& 9.5 \\
2		& 0.154		& 0.453	& \textit{3.60}	& 7.4		& 1.3		& 5.7 \\
3		& 0.472		& 1.43	& \textit{3.60}	& 8.2		& 3.9		& 9.2 \\
4		& 0.294		& 0.868	& \textit{3.60}	& 13.0	& 2.4		& 5.7 \\
\textbf{5}	& \begin{tabular}[t]{@{}l@{}}\textbf{0.567}\\0.483\end{tabular}	& \begin{tabular}[t]{@{}l@{}}\textbf{1.67}\\1.47\end{tabular}	& \textbf{3.60}	& \textbf{6.9}	& \textbf{4.3}	& \textbf{8.3} \\
6		& 2.24		& 6.61	& \textit{3.60}	& 12.1	& 18.2	& 19.9 \\
7		& 0.141		& 0.419	& \textit{3.60}	& 6.6		& 1.2		& 5.6 \\
8		& 0.189		& 0.566	& \textit{3.60}	& 7.1		& 1.6		& 7.8 \\
9		& 1.82		& 5.52	& \textit{3.60}	& 8.1		& 15.0	& 15.2 \\
10	& 0.265		& 0.811	& \textit{1.83}	& 7.1		& 2.7		& 7.7 \\
11	& 0.0342	& 0.122	& \textit{1.83}	& 4.1		& 0.4		& 3.0 \\
12	& 0.258		& 0.758	& \textit{1.83}	& 3.5		& 2.6		& 5.6 \\
13	& 0.305		& 0.888	& \textit{1.83}	& 5.4		& 3.1		& 9.3 \\
\textbf{14}	& \begin{tabular}[t]{@{}l@{}}\textbf{0.0714}\\0.0506\end{tabular}	& \begin{tabular}[t]{@{}l@{}}\textbf{0.210}\\0.145\end{tabular}	& \textbf{1.83}	& \textbf{3.6}		& \textbf{0.7}		& \textbf{4.6} \\
15	& 0.0391	& 0.114	& \textit{1.83}	& 3.5		& 0.4		& 6.6 \\
16	& 0.665		& 1.95	& \textit{1.83}	& 6.9		& 6.7		& 12.6 \\
17	& \begin{tabular}[t]{@{}l@{}}8.78\\7.88\\12.4\\9.95\\9.68\end{tabular}		& \begin{tabular}[t]{@{}l@{}}26.1\\23.7\\37.4\\29.3\\29.0\end{tabular}	& \textit{1.83}	& 41.4	& 98.0	& 89.7 \\
18	& \begin{tabular}[t]{@{}l@{}}9.37\\2.17\\3.95\\4.42\\2.41\end{tabular}		& \begin{tabular}[t]{@{}l@{}}27.8\\6.34\\11.8\\13.3\\7.22\end{tabular}	& \textit{1.83}	& 27.6	& 44.9	& 38.8 \\ \hline
\end{tabular}
\end{table*}

The Ottozawa area was used to study the effect of spatial variations in the radiocesium distribution on air dose rates. The range of dose rates measured at Ottozawa in July 2014 varied between \SIrange{3.5}{41.4}{\micro \sievert \per \hour} (Table~\ref{tab:ottozawa}). This is a higher range of values than measured in the five soil sampling campaigns (section~\ref{ssec:undisturbed2}), as Ottozawa lies closer to FDNPP than the sites in the five soil sampling campaigns and is more highly contaminated with fallout from the accident.

\begin{figure}
\centering
\includegraphics[width=0.4\textwidth]{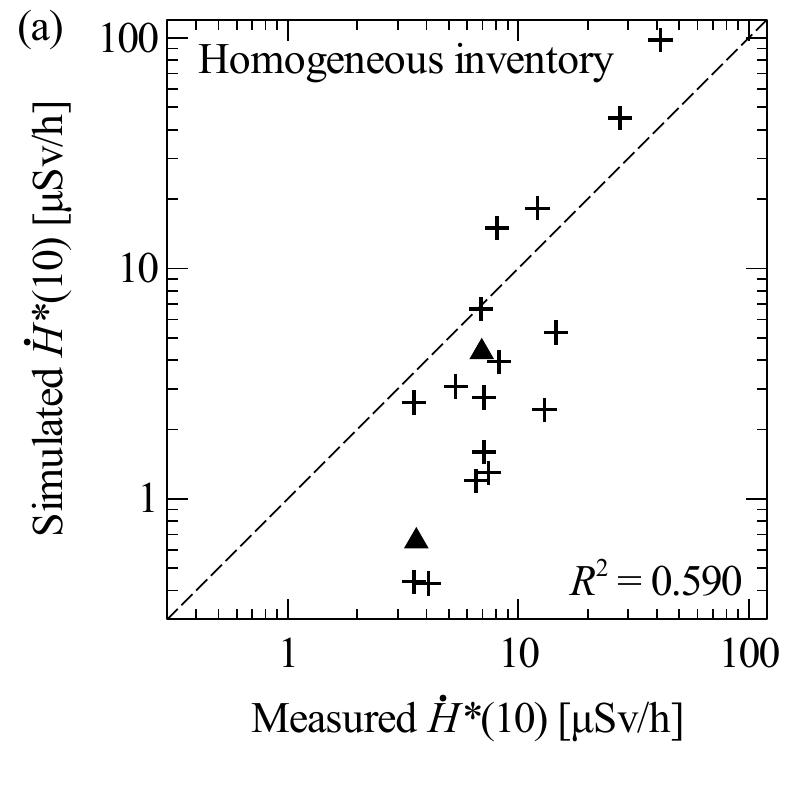} \includegraphics[width=0.4\textwidth]{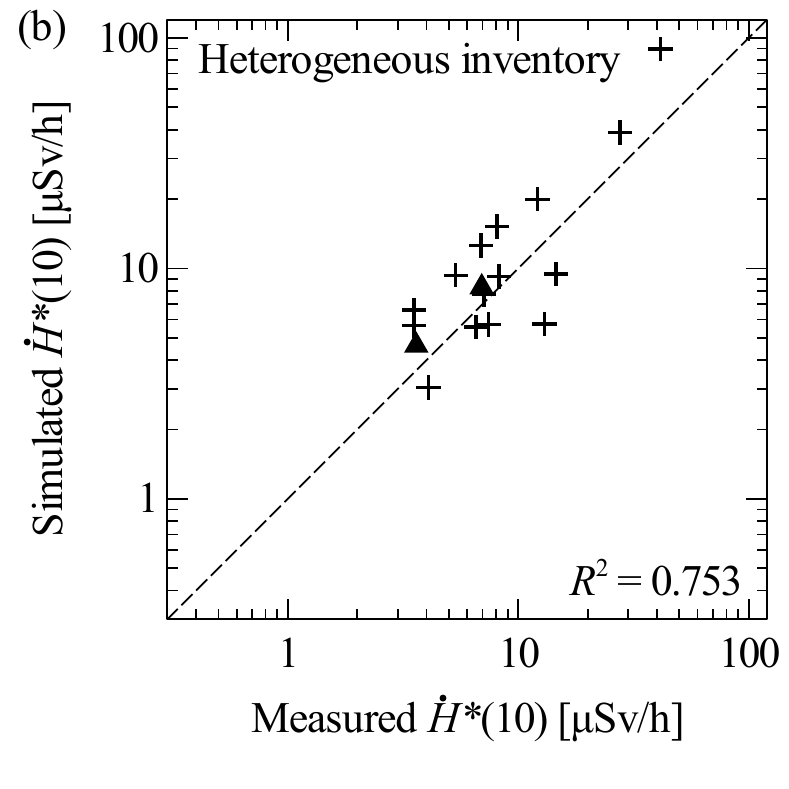}
\caption{\label{fig:ottozawa_scatter}Correlation between measured air dose rates and predictions from soil activity levels at Ottozawa. (a) Assumed a spatially homogeneous \textsuperscript{134}Cs and  \textsuperscript{137}Cs inventory. (b) Spatially varying inventory informed by all the soil sampling locations. Triangles indicate locations \num{5} and \num{14}, where scraper plate samples yielded the depth distribution.}
\end{figure}

Fig.~\ref{fig:ottozawa_scatter} shows two sets of predictions for the air dose rates from soil activity measurements, plotted against the dose rates measured in the field. The predictions shown in Fig.~\ref{fig:ottozawa_scatter}(a) did not account for the spatial variations in soil activity levels. Figure \ref{fig:ottozawa_scatter}(b) shows predictions from the models incorporating the measured spatial variations in the radiocesium inventory.

It is clear that modelling the spatial variations in the contamination distribution yields better predictions for the air dose rate. Therefore modelling the spatial variation is the better strategy if multiple soil samples across an area are available to include in the dose rate analysis.

The coefficient of determination is higher for the predictions modelling the spatial distribution ($R^2=0.753$) than for the predictions assuming a homogeneous radiocesium distribution ($R^2=0.590$). The origin of the difference in the $R^2$ values is traceable to the model accounting for the spatial distribution yielding a better prediction for the highest dose rate site, location \num{17}, with a measured dose rate of \SI{41.4}{\micro \sievert \per \hour}, than the model assuming a homogeneous cesium distribution.

The mean absolute percentage error for the predictions taking into account the spatial heterogeneity of the activity is \SI{47}{\percent}. This result is higher than the \SI{\approx 30}{\percent} mean absolute percentage error for the predictions for dose rates above flat, undisturbed fields (section~\ref{ssec:undisturbed2}). This difference is also observable by comparing the quality of the correlation in Figs.~\ref{fig:scatter_exact}(a) and \ref{fig:scatter_fit}(a) with Fig.~\ref{fig:ottozawa_scatter}(b).

\begin{figure}
\centering
\includegraphics[width=0.45\textwidth]{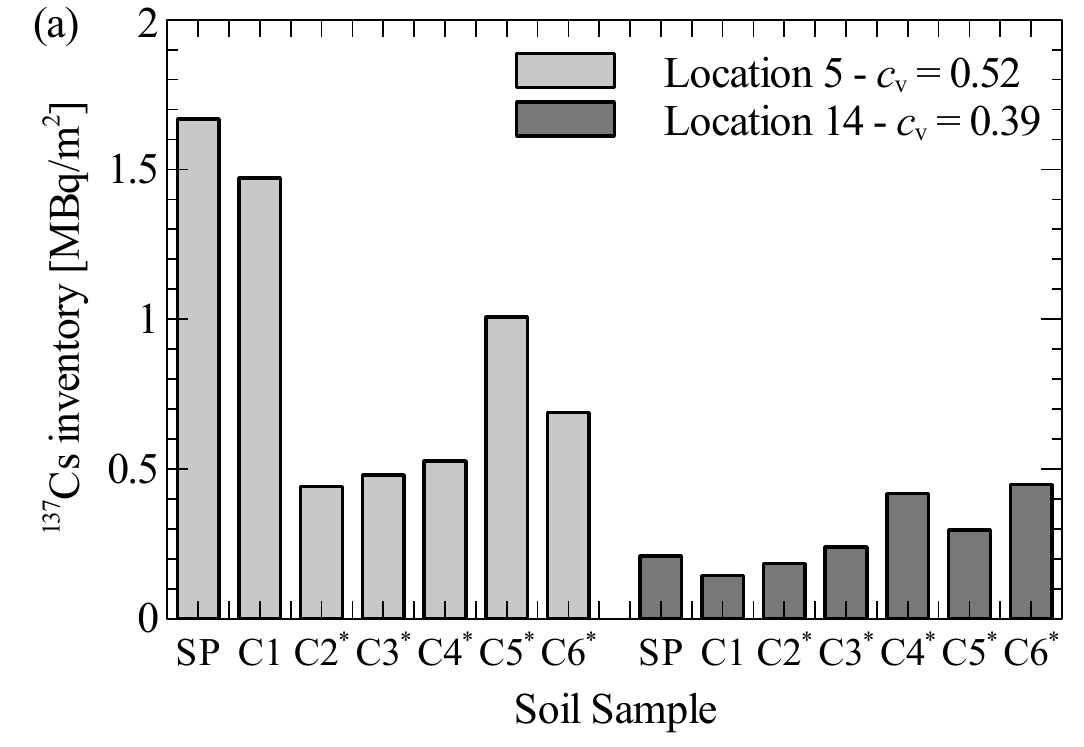} \includegraphics[width=0.45\textwidth]{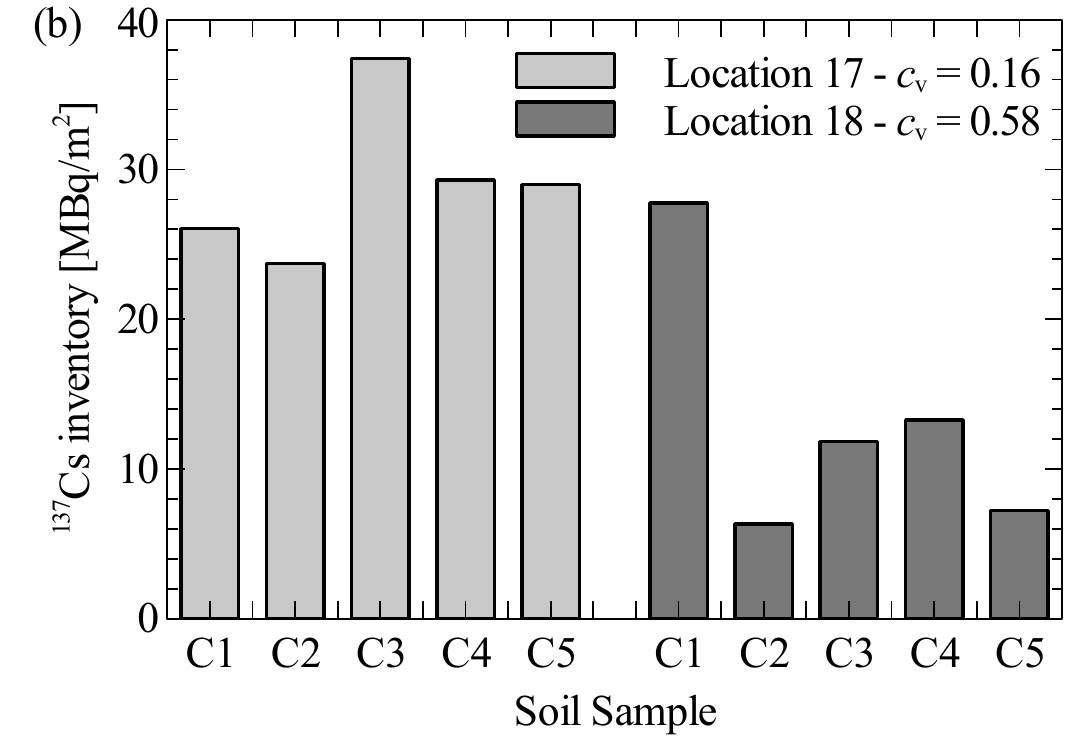}
\caption{\label{fig:ottozawa_variation}Variation in \textsuperscript{137}Cs inventory between soil samples at Ottozawa locations \numlist{5;14;17;18}. SP denotes a scraper plate sample, and C1, C2, etc.~denote the U-8 cup samples. Asterisks denote soil samples taken on October 10, 2014 and inventory decay corrected to July 24, 2014. $c_\mathrm{v}$ denotes coefficient of variation between the soil sample inventories at each location. $c_\mathrm{v}$ is the ratio of the sample standard deviation to the mean.}
\end{figure}

There are a number of distinctions between the modelling at Ottozawa and the flat, undisturbed fields that contribute to a higher uncertainty for the predictions at Ottozawa. There is a higher degree of measurement uncertainty for many of the soil samples at Ottozawa than for the sites visited in the five soil sampling campaigns, as the U-8 cup samples collect smaller volumes of soil than the scraper plate. This can be shown by examining the inventories from the multiple soil samples taken at locations \numlist{5;14;17;18} (Fig.~\ref{fig:ottozawa_variation}). There is large variation between the inventories between the samples at each location. The highest variation is seen for location \num{18}, where the largest inventory is four times greater than the smallest inventory.

The mean coefficient of variation for the four Ottozawa locations is $c_\mathrm{v}=0.41$. This is larger than the mean $c_\mathrm{v}=0.36$ observed by \citet{Saito2015a} for locations within a \SI{100}{\kilo \metre} radius of FDNPP, which are similar to the sites visited in the soil sampling campaigns. \citet{Mishra2015} independently reported a coefficient of variation of \num{0.27} between four samples at another site similar to those visited in the \citet{Matsuda2015} soil sampling campaigns.

Another factor contributing to the uncertainty for the predictions at Ottozawa include the fact that scraper plate samples were only taken at locations \num{5} and \num{14}. The depth distribution at other locations had to be inferred from these two measurements. It is notable that some of the best dose rate predictions obtained for Ottozawa were at locations \num{5} and \num{14} (triangles - Fig.~\ref{fig:ottozawa_scatter}(b)).

\subsection{\label{ssec:remediation2}Evaluation of farmland soil remediation methods}

\begin{table*}
\caption{\label{tab:remediation}The percentage reduction in the air dose rate after remediation of farmland soils by three different methods. The simulation input data (depth profiles, activity levels, etc.) were applicable on December 01, 2011. Full remediation means that all \num{149} by \num{149} cells on the simulation mesh were modelled as remediated land.}
\centering
\begin{tabular}{ l l l l l }\hline
Remediation	& Observed	& \multicolumn{3}{l}{Simulation results}\\ \cline{3-5}
method &  \begin{tabular}[t]{@{}l@{}}results\\\citep{JAEA2015b}\end{tabular} & \begin{tabular}[t]{@{}l@{}}Center of \num{37.5} by \SI{37.5}{\metre}\\remediated area\end{tabular} & 
\begin{tabular}[t]{@{}l@{}}Corner of \num{37.5} by \SI{37.5}{\metre}\\remediated area\end{tabular} &	
\begin{tabular}[t]{@{}l@{}}Full\\remediation\end{tabular} \\ \hline
Topsoil removal	& \SIrange{40}{70}{\percent}	& \SI{73}{\percent}	& \SI{65}{\percent}	& \SI{96}{\percent} \\
Reverse tillage	& \SIrange{30}{60}{\percent}	& \SI{54}{\percent}	& \SI{46}{\percent}	& \SI{71}{\percent} \\
\begin{tabular}[t]{@{}l@{}}Topsoil-subsoil\\layer interchange\end{tabular}	& \SI{\approx 65}{\percent}	& \SI{68}{\percent}	& \SI{60}{\percent}	& \SI{90}{\percent} \\ \hline
\end{tabular}
\end{table*}

We used the tool to evaluate the effectiveness of three methods for remediating farmland soils for decreasing air dose rates (Table~\ref{tab:remediation}). We calculated the reduction in air dose rate at the center and the corner of a \num{37.5} by \SI{37.5}{\metre} square area of remediated land, and compared with field results from a decontamination pilot project in Fukushima Prefecture \citep{JAEA2015b}. Also shown in Table~\ref{tab:remediation} are theoretical limits for the reduction in dose rates, calculated assuming remediation of all the land surface.

The performance of the topsoil removal and layer interchange methods of remediation are similar. Both methods yield \SI{\approx 65}{\percent} reduction in air dose rates for the square area of remediated land. These methods are more effective than reverse tillage, where the calculations indicated a \SI{\approx 50}{\percent} reduction in the air dose rate.

Experience from the decontamination pilot project \citep{JAEA2015b} suggests a range of dose rate reductions for topsoil removal and reverse tillage. A number of factors affect the percentage reduction in dose rates after land remediation, including the size of the area remediated, the homogeneity of the remediation actions, and the magnitude of the dose rate before remediation relative to the natural background dose rate. The remediation parameters, e.g.~the thickness of topsoil removed, or the depth of ploughing when performing reverse tillage, may also have varied slightly. The general correspondence in Table~\ref{tab:remediation} between the predictions from the tool and observed results is encouraging. However, the outlined uncertainties and the criticisms raised by \citet{Hardie2014} about the quantification of errors in the decontamination pilot project should be borne in mind when interpreting this conclusion.

One advantage of the layer interchange method over topsoil removal is that it does not create waste radioactive soil for disposal. However, the fact that the contaminated soil remains at the site after remediation, albeit below the ground surface, is tempered by the possible availability of the radioactive contaminants for uptake by crops or vegetation in future. This point may affect the viability of farming these lands after remediation if the crops or livestock produced approach food safety limits for radioactive cesium content.

\section{\label{sec:conclusions}Conclusions}

The simulation predictions for dose rates at flat, undisturbed fields from soil activity depth profiles showed good correlation with measurements. Little error was introduced by modelling exponential and hyperbolic secant fits to measured activity depth profiles. This conclusion necessarily depends on the experimental parameters for measuring activity depth distributions. Soil layers at least as fine as collected by \citet{Matsuda2015} are recommended if the data are to be used to evaluate air dose rates. Simulations of the Ottozawa area demonstrated that modelling spatial variations in contamination levels improves the quality of dose rate predictions. This approach is recommended if multiple soil activity samples across an area are available.

The main uncertainty in air dose rate predictions derived from soil samples is due to the sampling uncertainty for the true soil inventory distribution based on the limited volume samples. In situ or mobile gamma spectroscopy surveys offer a more comprehensive route to assess environmental radiocesium distributions, as they are subject to much lower sampling uncertainty~\citep{ICRU53}. The results from these surveys could be used to inform inputs for dose rate modelling and improve prediction quality. 

Simulations for the decrement in air dose rates seen at undisturbed, flat fields in Fukushima Prefecture for the first \num{20} months following the Fukushima Dai-ichi accident were consistent with the hypothesis that radiocesium decay and deeper migration in soil are the main responsible factors. Simulations of three farmland soil remediation methods for reducing air dose rates demonstrated that topsoil removal and layer interchange strategies have similar levels of effectiveness, and both methods are more effective than reverse tillage.

Techniques for modelling air dose rates from soil activity concentrations, such as described in this paper, would be effective for evaluating air dose rates in future and for planning land remediation works.

\begin{acknowledgments}

The decontamination pilot project was funded by the Cabinet and the Ministry of Environment. The authors are grateful to the town of Okuma for support of these investigations. We thank Satoshi Mikami for providing the mean air dose rates at flat, undisturbed fields from the air dose rate survey campaigns. We thank Kimiaki Saito for comments on the manuscript. We also thank colleagues within JAEA and Alan Cresswell for helpful discussions during the course of the research. Simulations were performed on JAEA's BX900 supercomputer.

\end{acknowledgments}

\bibliography{tool_paper}

\end{document}